 \documentclass[12pt,preprint]{aastex}

\newbox\grsign \setbox\grsign=\hbox{$>$} \newdimen\grdimen \grdimen=\ht\grsign
\newbox\simlessbox \newbox\simgreatbox
\setbox\simgreatbox=\hbox{\raise.5ex\hbox{$>$}\llap
     {\lower.5ex\hbox{$\sim$}}}\ht1=\grdimen\dp1=0pt
\setbox\simlessbox=\hbox{\raise.5ex\hbox{$<$}\llap
     {\lower.5ex\hbox{$\sim$}}}\ht2=\grdimen\dp2=0pt
\def\simgreat{\mathrel{\copy\simgreatbox}}
\def\simless{\mathrel{\copy\simlessbox}}
\newbox\simppropto
\setbox\simppropto=\hbox{\raise.5ex\hbox{$\sim$}\llap
     {\lower.5ex\hbox{$\propto$}}}\ht2=\grdimen\dp2=0pt

\slugcomment{To Appear in The Astrophysical Journal}

\shorttitle{Population Synthesis in the Blue II}
\shortauthors{Schiavon et al.}

\begin{document}


\title{Population Synthesis in the Blue II. The Spectroscopic Age
of 47 Tucanae}


\author{Ricardo P. Schiavon,  S. M. Faber}
\affil{UCO/Lick Observatory, University of California, Santa Cruz, CA 95064.}
\email{ripisc,faber@ucolick.org}

\author{James A. Rose}
\affil{Department of Physics and Astronomy, CB 3255, University of North
Carolina, Chapel Hill, NC 27599}
\email{jim@physics.unc.edu}

\and
\author{Bruno V. Castilho}
\affil{Laborat\'orio Nacional de Astrof\'\i sica, MCT
CP 21, 37500-000 Itajub\'a, Brazil}
\email{bruno@lna.br}


%\altaffiltext{1}{Visiting Astronomer, Cerro Tololo Inter-American Observatory.
%CTIO is operated by AURA, Inc.\ under contract to the National Science
%\altaffiltext{2}{Society of Fellows, Harvard University.}
%\altaffiltext{3}{present address: Center for Astrophysics,
%    60 Garden Street, Cambridge, MA 02138}
%\altaffiltext{4}{Visiting Programmer, Space Telescope Science Institute}
%\altaffiltext{5}{Patron, Alonso's Bar and Grill}


\begin{abstract}

We develop a new set of models for intermediate-metallicity single
stellar populations in the blue/optical region and use those models
to determine the spectroscopic age of 47 Tuc. The models are based
on a moderately high-resolution (1.8 {\AA} FWHM) empirical spectral
library, state-of-the-art theoretical isochrones from M. Salaris and
the most recent set from the Padova group, and new semi-empirical
calibrations between fundamental stellar parameters and observables.
Model line-strengths include all corrections for deficiencies of
the stellar library that are described in Paper I.  We highlight
the importance of correctly modeling the luminosity function (LF)
of the cluster at the level of the giant branch, in order to achieve
a good reproduction of the integrated spectrum; agreement between the
spectroscopic age and the age based on the cluster's color-magnitude
diagram (CMD) is achieved only if the {\it observed} LF is used rather
than the theoretical ones, which either do not include AGB stars
(Salaris) or underpredict the total number counts of bright giants in
the cluster by a factor of two (Padova). After all corrections are made,
the CMD and the spectroscopic ages (from $H\gamma$ and $H\beta$) are in
close agreement: $\sim$ 11--12 Gyrs for Salaris isochrones and $\sim$
13 Gyrs for Padova. The difference between the model ages is due to
the inclusion of atomic diffusion in the Salaris models. Previously
older spectroscopic ages were due to the underestimate of the number of
red giants and/or the use of isochrones that neglected the effects of
He-diffusion and $\alpha$-enhancement. Uncertainties in spectroscopic age
determinations of old stellar populations stem from a number of effects,
the most important of which are the $T_{eff}$ and {\it [Fe/H]}-scales of
the giant stars used in the stellar library, the LF on the upper giant
branch, and the assumed metallicity of the target stellar population
itself. A $\pm$ 1 Gyr uncertainty in age results from uncertainties of
$\pm$ 75 K in the $T_{eff}$-scale of the library giants, $\pm$ 0.1 dex in
the level of the giant-branch LF, and $\pm$ 0.1 dex in the assumed {\it
[Fe/H]} of either the target stellar population or the assumed zeropoint
of the metallicity scale of the stellar library. A similar underestimate
in the bright giant LF, if it exists in current super-solar metallicity
models, would cause spectroscopic ages of elliptical galaxies inferred
from such models to be too high by approximately 30\%.

\end{abstract}




\section{Introduction}

In Paper I of this series (Schiavon et al. 2002), we performed
an empirical synthesis of the integrated spectrum of the Galactic
metal-rich globular cluster 47 Tucanae. The model spectrum was computed
directly from a color-magnitude diagram (CMD) of the cluster, being thus
essentially independent of inputs from theoretical isochrones.  In that
paper we achieved good agreement between models and observations in the
equivalent widths of two Balmer lines and several metal lines. In order
to reach such agreement, we needed to correct the model line indices for
two limitations of the spectral library adopted in our models, namely,
the relative paucity of sufficiently metal-poor giants and the absence of
CN-strong stars, which are known to be present in the cluster.  We also
had to require that the iron abundance of 47 Tuc be {\it [Fe/H]}=--0.75,
which is slightly below and yet within the errorbars of the standard
value determined by Carretta \& Gratton (1997) ([Fe/H]=--0.7).

In this second paper of the series, we turn to the more challenging task
of making models based on theoretical isochrones.  All the corrections
described above, which were inferred in Paper I, are applied to the
models computed here. Moreover, we subtract from the observed indices the
contribution by blue stragglers, which are not included in the theoretical
isochrones. It was shown in Paper I that such corrections are important
only for $H\delta_F$ and, to a lesser degree, $H\gamma_{\sigma<130}$. The
effect of blue stragglers on all indices studied in this paper is shown
in Table \ref{tbl-4}. The index values adopted in this paper are those
listed in the second row of Table 3 of Paper I.

The resultant model spectra computed from the theoretical isochrones are
used to deduce the cluster's age. Such a test is of fundamental importance
for stellar population models. A key requirement is that the same age has
to be obtained from the fit of the theoretical isochrone to the cluster's
CMD, and from the fit of theoretical spectra to the integrated spectrum
of the cluster (the so-called spectroscopic age). This is especially
interesting in view of the exceedingly high spectroscopic age ($>$ 20
Gyrs) that was obtained for 47 Tuc from the equivalent width of $H\gamma$
by Gibson et al. (1999). The mismatch between spectroscopic and CMD-based
ages may even be more general, because Cohen, Blakeslee \& Rhyzov (1998)
have found very high spectroscopic ages for a sample of Galactic GCs, on
the basis of the equivalent width of $H\beta$. More recently, Vazdekis
et al. (2001) showed that a lower spectroscopic age ($\sim$ 13 Gyrs)
is obtained when isochrones that take into consideration effects such as
$\alpha$-enhancement and diffusion of heavy elements are used. However,
Vazdekis et al. still obtained a residual difference of $\sim$ 4 Gyrs
between their spectroscopic and CMD-based ages, in the sense that the
spectroscopic age was higher.

In the present paper, we approach this problem as follows. We first infer
an age for the cluster by fitting theoretical isochrones to its CMD. We
adopt the isochrones by Salaris and collaborators, which are the same
as those employed by Vazdekis et al. (2001). For comparison purposes,
we perform the same exercise adopting $\alpha$-enhanced isochrones by
Salasnich et al. (2000, hereafter, Padova isochrones). The isochrones
are compared to cluster data in both the CMD and luminosity function
(LF) domains. A major conclusion is that the raw isochrones adopted do
not reproduce the cluster's LF correctly. They underestimate the number
counts of giants brighter than the horizontal branch by roughly a factor
of 2.5, in part because the theoretical isochrones do not include AGB
stars. When we adopt the Padova isochrones, which do include AGB stars,
the mismatch is reduced by $\sim$ 0.1 dex. We then infer corrections to
the model predictions by forcing the theoretical LF to match the observed
one. Such corrections are shown to be very important, implying a decrease
of the order of 3--4 Gyrs in the spectroscopic age inferred from Balmer
lines.  The LF of giant stars is thus a key input in deriving accurate
spectroscopic ages of old stellar populations. The revised spectroscopic
ages inferred when adopting the ``LF-corrected'' isochrones are in
excellent agreement with those obtained from the fits to the CMD. Via
this careful two-step procedure in Papers I and II, we now feel that
the mysterious spectroscopic age discrepancy for 47 Tuc has been resolved.

The paper is organized as follows. In Section \ref{isocmd} the theoretical
isochrones are compared to data on 47 Tuc in both the CMD and LF domains.
In Section \ref{isoobs} we compare observed line-indices to model
predictions computed from both the original and the ``LF-corrected''
isochrones.  Model uncertainties due to errors in the input stellar
parameters are discussed in Section \ref{unc}, and our conclusions are
summarized in Section \ref{conc}.

\section{Theoretical Isochrones vs. Data} \label{isocmd}

\subsection{Color-Magnitude Diagram} \label{cmd}

\begin{figure}
\plotone{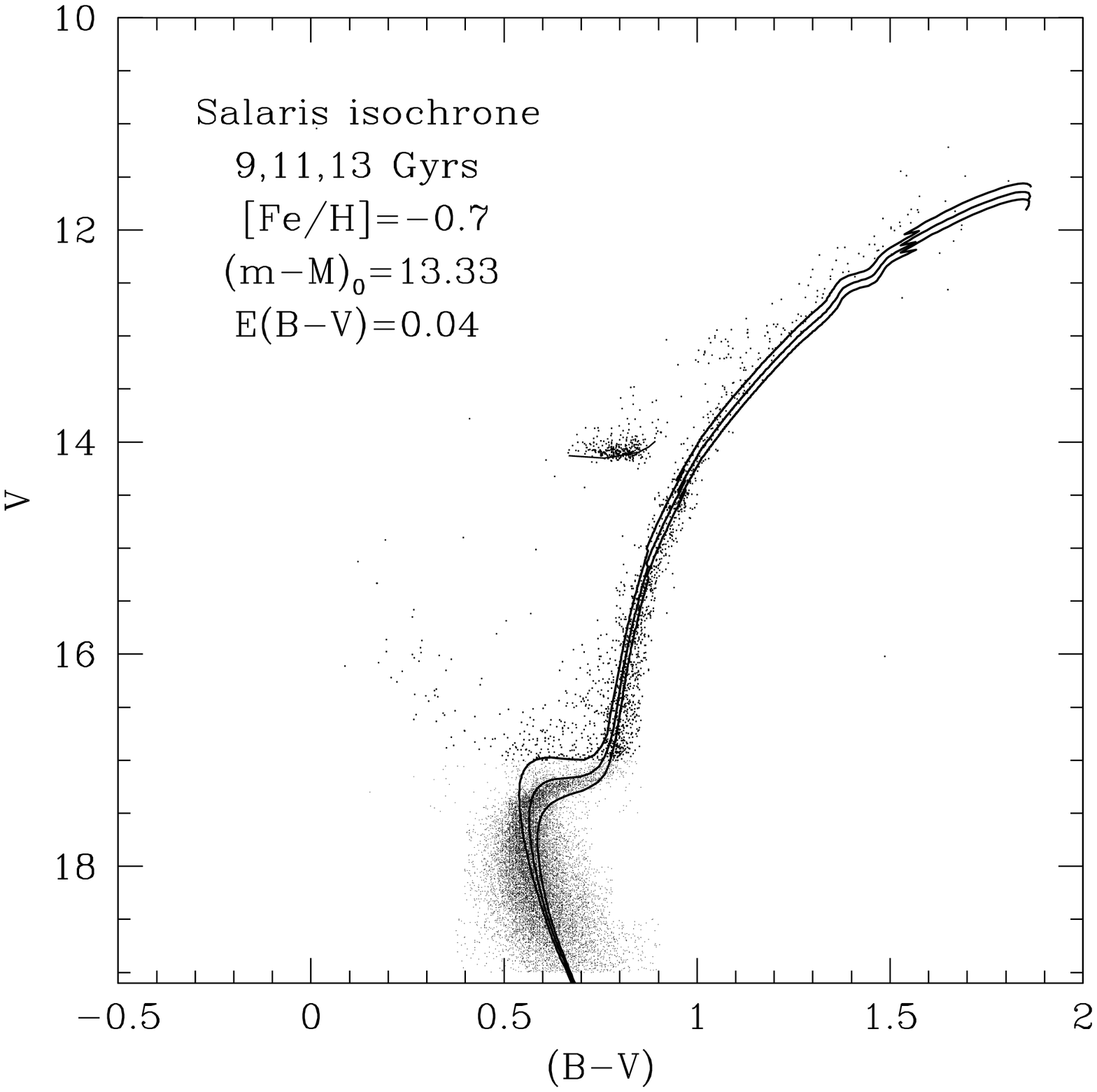}
\caption{CMD of 47 Tuc out to a radius of $\sim$ 30'' from Howell et
al. (2000). The theoretical isochrones are described in Vazdekis et
al. (2001). They have been converted to the observational plane by
adopting the calibrations of Alonso et al. (1995, 1996) for dwarfs and
the calibration presented in Paper I for the giants.  Adopted values for
the distance modulus and {\it E(B--V)} are displayed in the upper left
corner and were taken from Kaluzny et al. (1998). The best-fitting age
according to those isochrones is 11--12 Gyrs.
\label{fig1}
}
\end{figure}

Before comparing the isochrone-based model spectra with the observations,
we call attention to important features of the theoretical isochrones
that are crucial to interpreting our results. In Figure \ref{fig1}
the CMD of 47 Tuc (Howell, Guhathakurta \& Gilliland 2000) is overlaid
with isochrones from Salaris and collaborators for 8, 10, 12 and 14
Gyrs. Conversion from the theoretical plane ($L_{bol}$,$T_{eff}$) to
the observational plane [$V$,{\it (B--V)}] was performed according to
the recipe described in Paper I, and adopting a distance modulus of
$(m-M)_0$=13.33 and {\it E(B--V)}=0.04 (Kaluzny et al. 1998). These
isochrones are described in Vazdekis et al. (2001) and were kindly
provided by M. Salaris. They are computed taking into consideration
diffusion of heavy elements and an $\alpha$-enhanced mixture ({\it
[Fe/H]}=--0.7, {\it [$\alpha$/Fe]}=+0.4). These were the isochrones
used by Vazdekis et al. (2001) to infer a spectroscopic age of
$\sim$ 14 Gyrs for 47 Tuc, on the basis of the measurement of the
$H\gamma_{\sigma<130}$ index (for a definition of that index, see Figure
16 of Paper I). The age inferred from their fit to the CMD of the cluster
is significantly younger: $\sim$ 10 Gyrs. In spite of this discrepancy,
Vazdekis et al. (2001) greatly improved over previous spectroscopic
age determinations, which gave ages in excess of 20 Gyrs (Gibson et
al. 1999). In particular, they showed that $\alpha$-enhancement and
diffusion of heavy elements cause a reduction in both the CMD and
spectroscopic ages of 47 Tuc by several Gyrs.

From Figure \ref{fig1} it can be seen that the best match to the position
of the cluster's turn-off in the CMD is achieved for an isochrone with age
11-12 Gyrs. This is of course dependent on the reddening assumed for the
cluster. An error of 0.01 mag in {\it E(B--V)} results in an uncertainty
of $\sim$ 1 Gyr in the age inferred from CMD-fitting. That is the reason
why we infer a slightly older age than Vazdekis et al., even though we
are using the same set of isochrones, as their adopted {\it E(B--V)}
is 0.01 mag higher than ours.  The isochrone for 11 Gyrs matches very
well the color of the turn-off of the cluster.  However, it predicts a
subgiant branch which is too bright by 0.05 mag. There also are mismatches
in the color of the giant branch. The model isochrone is too red by $\sim$
0.03 mag for giants brighter than V $\sim$ 14 and too blue by a comparable
amount at the base of the giant branch.  Adopting the Alonso et al. (1999)
calibration of $T_{eff}$s vs. $(B-V)$, rather than the recipe of Paper I,
leads to a slightly better match in the base of the giant branch, but it
worsens the match at $V \sim 14$, where our calibration provides a good
match to the data. Such mismatches, if due to errors in the $T_{eff}$s
of the isochrones, may have a non-negligible impact on the ages inferred
from the fit of the integrated spectrum of the cluster, as we show below.
It is also important to notice that AGB stars are not included in the
Salaris isochrones. It will be shown in Section \ref{isoobs} that this
also has an important impact on the integrated line indices.

\begin{figure}
\plotone{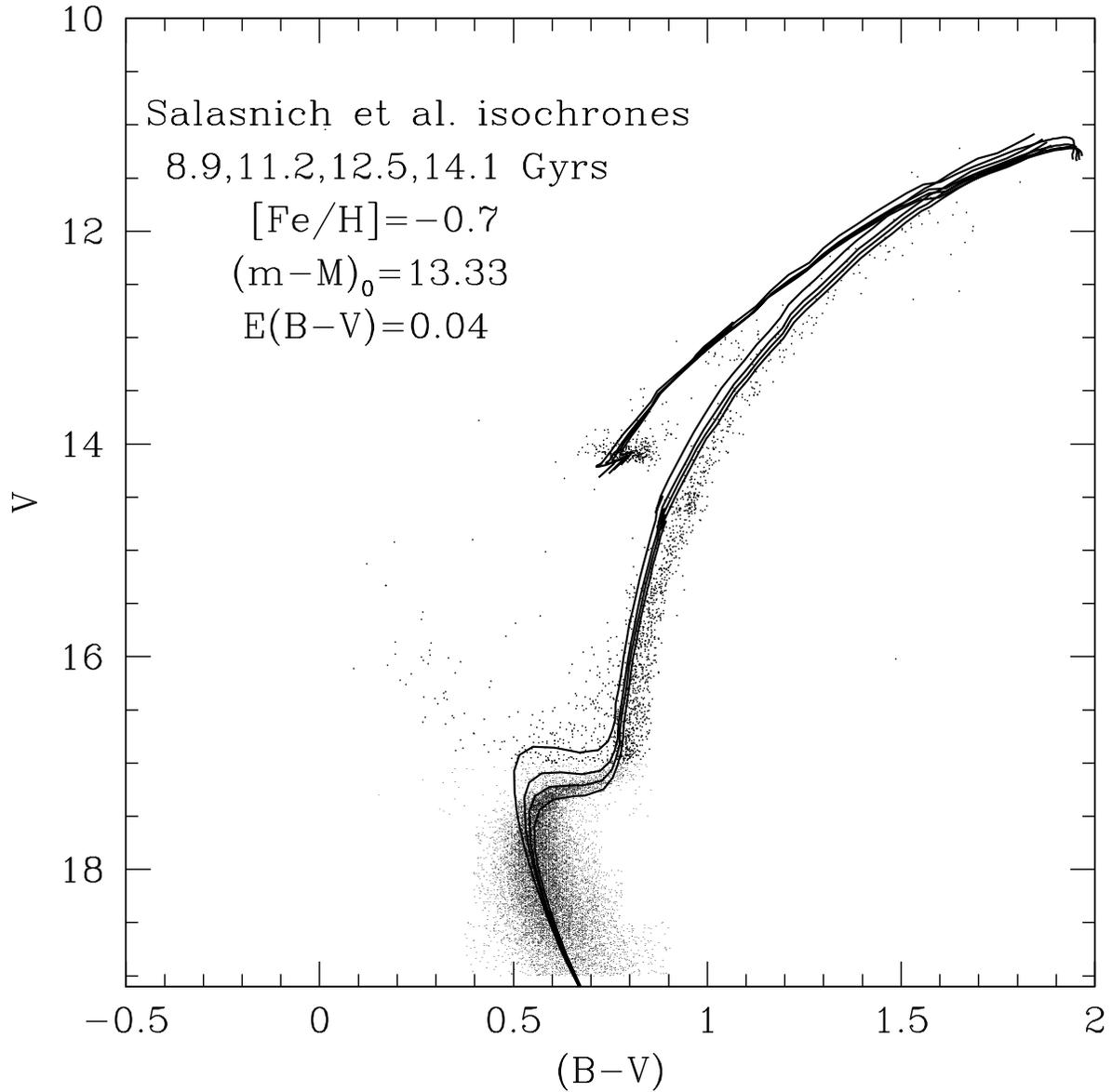}
\caption{Same as Figure \ref{fig1}, but now comparing the data on 47 Tuc
with the isochrones from Salasnich et al. (2000). Those isochrones were
computed using the same chemical composition of the Salaris isochrones
but do not include diffusion of heavy elements. For that reason, the age
that best fits the turn-off color and luminosity is $\sim$ 1 Gyr older
than in the case of the Salaris isochrone.
\label{fig2}
}
\end{figure}

In Figure \ref{fig2} we repeat the CMD of 47 Tuc, this time comparing
the data with the Padova isochrones.  Theoretical quantities ($T_{eff}$
and $M_{bol}$) were transformed into the observational plane following
the same recipes as adopted to the Salaris isochrones shown in
Figure \ref{fig1}.  The Padova isochrones were computed assuming
the same chemical composition as adopted in the computation of the
Salaris isochrones.  However, the Padova isochrones do not consider
the effect of diffusion of helium and heavier elements. As explained
by Vazdekis et al. (2001), inclusion of heavy-element diffusion results
in a slightly cooler and fainter turn-off for a given stellar mass and
chemical composition. In fact, the turn-off of the 11 Gyrs isochrone from
Salaris is about 0.16 mag fainter and 150 K cooler than the turn-off of
the Padova isochrone for a similar age (11.2 Gyrs). For this reason the
age that best fits the turn-off color and magnitude of 47 Tuc, according
to the Padova isochrones, is somewhere between 12.5 and 14.1 Gyrs, being
thus slightly older than the one inferred from the Salaris isochrones
(Figure \ref{fig1}).

Evidence for the occurence of atomic diffusion in the Sun has been
presented in a number of studies. For instance, Basu, Pinsonneault \&
Bahcall (2000) show that diffusion of heavy elements is required if models
are to fit the data on the profiles of sound speed, density and adiabatic
index in the Sun. For more metal-poor stars, Lebreton et al. (1999)
showed that the HR diagram from Hipparcos data for stars with --1.05 $<$
{\it [Fe/H]} $<$ --0.45 can be reconciled with the predictions of stellar
evolution only if diffusion of heavy elements is considered. However, the
extent to which atomic diffusion affects stellar structure and evolution
is still under debate. It has been argued that the constancy of the Li
abundance in metal-poor dwarfs with a wide range of metallicities (the
so-called lithium plateau, Spite \& Spite 1982) suggests that diffusion
effects may be partially inhibited (Deliyannis \& Demarque 1991).
On the other hand, Salaris \& Weiss (2001) showed that models with fully
efficient diffusion may be consistent with the lithium plateau if the
errors on lithium abundances and sample incompleteness are duly accounted
for. According to Vazdekis et al. (2001), the consideration of atomic
diffusion in its full extent reduces the turn-off age of 47 Tuc by $\sim$
1 Gyr. An important constraint on the extent of diffusion affects has also
been placed by Gratton et al. (2001), who found that {\it [Fe/H]} is the
same for stars at the turn-off and at the base of the giant branch of the
metal-poor globular cluster NGC 6397. Models by Chaboyer et al. (2001)
show that the latter result implies an inhibition of diffusion in the
outer layers of stellar models, such that the reduction of CMD-based ages
due to atomic diffusion is only 4\% ($\sim$ 0.5 Gyr in the case of 47
Tuc). In summary, though we tend to favor the age based on Salaris models
(11 Gyrs), the main emphasis of this work is on making sure that the
spectroscopic age we obtain is {\it consistent} with the one based on the
fit of the cluster turn-off, for whatever set of isochrones is employed.

In Figure \ref{fig2} it also can be seen that the Padova isochrones
do not perfectly fit the color of the giant branch, the mismatch being
similar (if slightly worse) to the one mentioned above for the Salaris
isochrones: Padova isochrones are bluer than the observations by $\sim$
0.04 mag in the giant branch. We note that the Padova isochrones do
include the AGB phase, but it is also displaced to the blue, as compared
to the observations, by the same amount as the giant branch. The position
of the HB is well matched by the Padova isochrones, but its extension
is underpredicted.

\subsection{Luminosity Function} \label{lfun}

\begin{figure}
\plotone{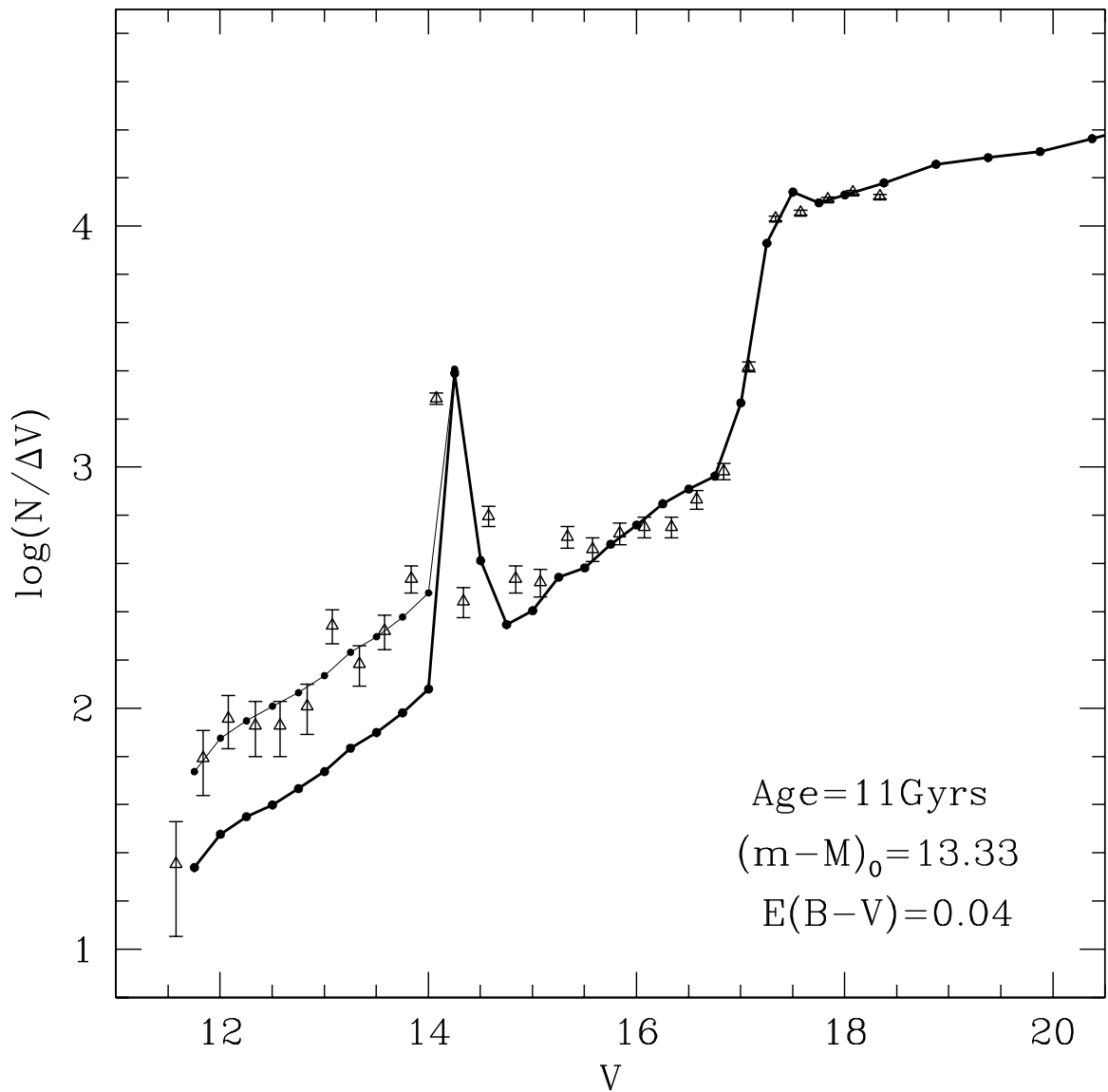}
\caption{Comparison between the observed luminosity function of 47 Tuc
and that obtained from the best fitting Salaris isochrone. The thick
line represents the original Salaris LF, which lacks AGB stars. The thin
line represents our ``corrected'' LF, in which a constant factor of 0.4
dex has been applied to the number counts brighter than V $\sim$ 14 in
order to bring them into accordance with the observed LF. The secondary
peak in the observational data at V $\sim$ 13 is the AGB bump.
\label{fig3}
}
\end{figure}

Matching a CMD such as the one shown in Figure \ref{fig1} is just the
first step when comparing a given theoretical isochrone with cluster data;
matching the LF of the cluster is also extremely important. In fact,
the LF of giant stars has a large and not often appreciated impact on
the integrated spectrum in general, and on line indices in particular. In
Figure \ref{fig3}, we compare the LF measured from the data with the LF
of the first-ascent giant-branch predicted by the best-fitting Salaris
theoretical isochrone of age 11 Gyrs (thick line).  The observed LF
is in good agreement with data on 47 Tuc from other sources in the
literature, namely Hesser et al. (1987), Kaluzny et al. (1998), and
Zoccali et al. (2000) (the last being based roughly on the same data
as this work). The theoretical LF is a good match to the observed one
up to the horizontal branch. For brighter stars, however, the Salaris
LF underpredicts the observations by $\sim$ 0.4 dex on average (the
mismatch is slightly lower at $V \sim 12.5$ and higher at the location
of the AGB bump, at $V \sim 13$).

The LF mismatch is in part due to the fact that the Salaris LF does not
include AGB stars. However, on purely theoretical grounds it is hard to
attribute the mismatch solely to the lack of these stars, as current
stellar evolution models predict a ratio of $\sim$ 0.15 between the
lifetimes in the AGB and HB phases (Cassisi et al. 2001).  At the AGB
bump (V $\sim$ 13.1), the observed number of AGB stars is roughly the
same as first-ascent giants: in the interval 12.8 $< V <$ 13.2, there
are 30 AGB and 32 first-ascent RGB stars.  Thus, adding AGB stars would
imply roughly a factor of two (0.3 dex) increase, in the raw theoretical
LF at the AGB-bump level. However, the difference between the observed
and theoretical LFs at the level of the AGB bump amounts to $\sim$ 0.5
dex ($V \sim 13.1$ in Figure \ref{fig3}), which leaves an extra 0.2 dex
mismatch between the pure RGB observed LF and the theoretical one. For
brighter stars, it becomes increasingly difficult to distinguish between
AGB and first-ascent RGB stars, but it is doubtful that the ratio changes
very much. At fainter magnitudes, below the AGB bump, there are even
fewer observed AGB stars.  Therefore we conclude that only at most 2/3 of
the mismatch between the Salaris LF and the data at the level of the AGB
bump could be due to the lack of AGB stars in the theoretical isochrone,
and that fraction is probably lower in other regions of the AGB, where
AGB stars are even more sparsely distributed. In short, some increase
in the model prediction for the number of model {\it first-ascent}
giants above the HB seems also to be needed.

\begin{figure}
\plotone{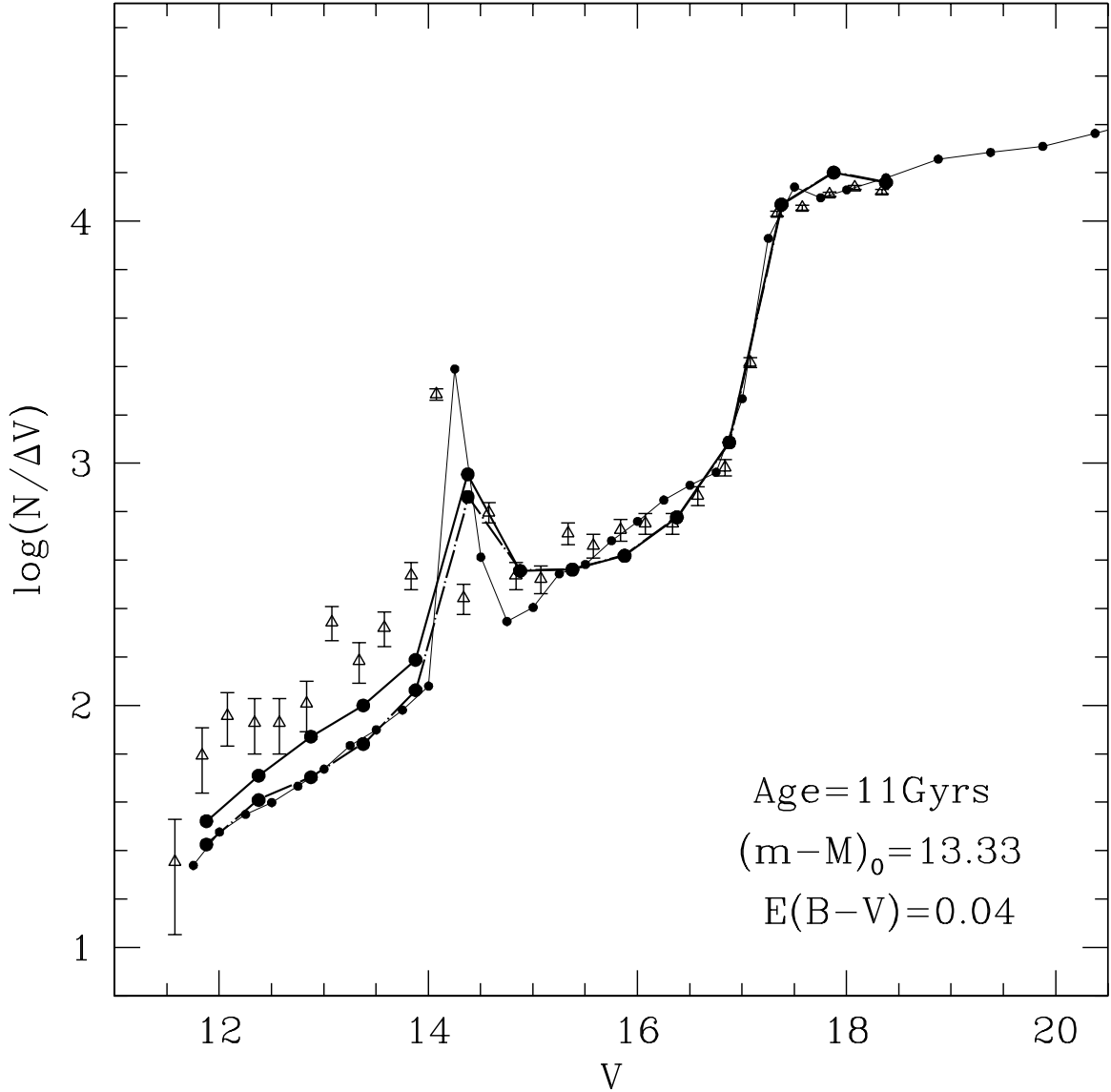}
\caption{The observed LF of 47 Tuc versus the LF from the Padova
isochrones (Salasnich et al. 2000). The thick solid line represents the
original LF, while the thick dash-dotted line represents the LF that
results when AGB stars are artificially removed. For reference, we repeat
from Figure \ref{fig3} the LF inferred from the Salaris isochrone (thin
line). The broader Horizontal Branch in the Padova isochrones (V $\sim$
14.5) is due to their coarser mass bins, but total number counts in the
Horizontal Branch are the same.  When stripped of AGB stars, the Padova
LF is in very good agreement with the one from Salaris. However, the
full Padova isochrone, which does include AGB stars, still underpredicts
the observed number counts above the Horizontal Branch by approximately
0.3 dex.
\label{fig4}
}
\end{figure}

This conclusion can be checked by comparing to the predicted LF
of the Padova isochrones, which include both first-ascent and AGB
stars explicitly. This comparison is shown in Figure \ref{fig4} which
illustrates theoretical LFs based on the Padova isochrones for 11.2
Gyrs. The thick solid line represents the total Padova LF, while the thick
dash-dotted line represents the same LF after removal of the AGB stars.
For reference, we also reproduce the original Salaris LF from Figure
\ref{fig3}, now as a thin line. We note that because the Padova isochrones
are provided in coarser mass steps, the signature of the Horizontal Branch
in the Padova LF is broader than in the case of the Salaris LF, but the
total number of HB stars is similar.  Two conclusions can be drawn from
this figure. First, we call attention to the very good agreement between
the two sets of pure RGB theoretical isochrones (thin line and thick
dashed line), which means that both groups predict similar lifetimes
for first-ascent giant stars. This has also been shown by Zoccali \&
Piotto (2000). Second, and more importantly, even when the AGB stars
in the Padova isochrones are considered (thick solid line), the number
counts above the Horizontal Branch are still underpredicted by $\sim$
0.2-0.3 dex. This confirms our suspicion above that the theoretical
number of first-ascent giants above the HB is also too smal. Evidently
the shortfall is similar in both the Salaris and Padova models. 

We recall that our observed LF is in good agreement with that of other
works, like Hesser et al. (1987), Kaluzny et al. (1998) and Zoccali
et al. (2000). In particular, the LFs of Hesser et al. and Kaluzny et
al. do not refer to the central parts of the cluster, so that it is
very unlikely that the LF mismatch is due to an environmental effect,
or that it may be afflicted by observational errors due to crowding.

Another feature that can affect the observed number ratio of giant to main
sequence stars is binarity. Throughout this discussion, we have adopted
the turn-off region ($V \sim 17.5$) as the normalization point to compare
the theoretical LF to the observations. However, if the number counts at
the level of the turn-off and upper main sequence are underestimated due
to the presence of binary stars, this could be responsible for part of the
LF mismatch of Figures \ref{fig3} and \ref{fig4}. The color distribution
of stars within narrow magnitude ranges of the main sequence of 47 Tuc
indeed are not symmetrical, presenting an extended tail towards redder
$(B-V)$, which is a well-known signature of binary contamination. In fact,
Albrow et al. (2001) have determined a binary fraction of 14\% $\pm$ 4\%
from observations of upper main sequence stars in the core of 47 Tuc.
In a study of binarity effects on the LF of NGC 6752, Rubenstein \&
Bailyn (1999) found that the turn-off luminosity function needs to be
corrected upwards by $\sim$ 0.1 dex, if the binary fraction is 22\%
at about 1 mag below the turn-off. We therefore conclude that, since
the binary fraction of 47 Tuc is lower than that of NGC 6752, at a
comparable position in the main sequence, the correction on the LF of
47 Tuc is lower than 0.1 dex, and thus is not able to account for the
mismatch seen in Figures \ref{fig3} and \ref{fig4}. Moreover, we note
that a correction for the effect of binaries, while bringing a slight
improvement to the match in the upper giant branch, would bring the LF
into disagreement in the lower giant branch.

The situation regarding the match of stellar evolution models to
the LFs of first-ascent giants in clusters is controversial. That
the theoretical models do not provide a perfect match to the number
counts of giants in some clusters has been pointed out some time ago
(Stetson 1991, Bergbusch \& Vandenberg 1992, Bolte 1994, Sandquist et
al. 1996, Sandquist et al. 1999). In a recent paper, Langer, Bolte \&
Sandquist (2000) found a comparable mismatch ({\it i.e.} $\sim$ 0.2
dex underestimate above the HB) between observed and theoretical LFs
of first-ascent RGB stars for the metal-poor clusters M5 and M30. In
particular, they note that the models underpredict the number of stars
at different levels of the giant branch for the two clusters. For M30,
the mismatch happens for giants below the RGB bump, whereas for M5 the
models underpredict the number of giants brighter than the RGB bump,
similarly to what we found for 47 Tuc.  Langer et al. suggest that the
mismatch can be due to deep mixing, which would bring fresh fuel to the
hydrogen-burning shell, thus prolonging the lifetimes of first-ascent
RGB stars. On the other hand, Zoccali \& Piotto (2000) compared the
observed LFs of a number of clusters over a broad range of metallicities
to theoretical models by different groups, including Padova, Straniero
et al. (1997), previous models by Salaris and collaborators (Salaris \&
Weiss 1998).  Contrary to the conclusions of Langer et al. (2000), they
found rather good agreement for all clusters more metal-poor than {\it
[Fe/H]} $\sim$ --0.7.  However, in their Figure 5 a slight mismatch
(up to $\simless$ 0.2 dex), in the same sense as found in this work,
can be seen in the LF of 47 Tuc for $V \simless 13.5$.

While there still is controversy concerning the ability of
theoretical models to predict the LF of first-ascent giants, the
AGB LF is largely unexplored. To our knowledge, there is as yet no
direct detailed comparison in the literature between number counts
of AGB stars in clusters and model LFs.  This is partly due to the
paucity of statistically representative samples of cluster AGB stars
in the literature, mostly on account of their short lifetimes. On the
theoretical side, AGB models are subject to important uncertainties that
may affect lifetime predictions (Cassisi et al. 2001), mostly related to
the unknown efficiency of macroscopic mechanisms, like convection and
diffusion, and to the input physics of the models (equation of state,
opacities and nuclear cross sections). On the calibration side, the
matter is further complicated by the uncertainties in the bolometric
corrections of M giants, especially for the most metal-rich clusters,
where molecular opacities in the V band, and in particular their
dependence on metallicity, are difficult to model (see, for instance,
Westera et al. 2002). 

Maraston (1998) compared model predictions, based on the fuel consumption
theorem, to the fractional contribution of RGB and AGB stars from NGC
6528 and NGC 6553 to the integrated light in the V band, finding good
agreement. This result is difficult to reconcile with our findings,
although Maraston's models are based on a different set of theoretical
models. Moreover, Maraston's test does not disentangle the effect of the
number of AGBs and their V-band luminosities, which are both subject
to significant uncertainties. Taken at face value, Maraston's result
suggests that there may not be a LF mismatch for clusters of near-solar
metallicity, although this is at odds with the finding, by Zoccali \&
Piotto (2000), that model LFs are a poor match to the number counts of
near-solar metallicity clusters in the upper giant branch.

Regardless of the origin of the LF discrepancy in Figure \ref{fig3}, its
presence must be considered in order to account correctly for
all evolutionary stages in the CMD of Figure \ref{fig1}.  We also want
to compare spectroscopic ages based on both the raw and corrected LFs,
in order to compare our results to those of Vazdekis et al. (2001),
who used the Salaris isochrones and (apparently) the associated raw
LF. This will be discussed in Section \ref{isoobs}.

\subsection{Effective Temperatures and Colors of the Red Giants} \label{tefcol}

Another important issue concerns the interplay between the $T_{eff}$-scale
of the model isochrones and the $T_{eff}$ $vs.$ {\it (B--V)} relation used
in the conversion of the theoretical isochrones to the observational
plane. As noted above, the Salaris isochrone that best matches the
cluster's turn-off ($\sim$ 11 Gyrs) presents a mismatch of as much as
$\sim$ 0.03 mag in the color of the red giant branch. However, even a
successful fit to the giant branch {\it (B--V)} color in the CMD of a
cluster would {\it not} mean that both the theoretical $T_{eff}$-scale and
the $T_{eff}$ $vs.$ {\it (B--V)} are individually correct.  As stressed by
Weiss \& Salaris (1999) and Salaris, Cassisi \& Weiss (2002), a suitable
{\it combination} of a wrong $T_{eff}$-scale with a wrong $T_{eff}$ $vs.$
color relation can produce a successful match to the giant branch. For
the same reason, the slight mismatch of 0.03 mag between the theoretical
isochrone for 11 Gyrs and the data on the giant branch of 47 Tuc could
be due to a wrong $T_{eff}$-scale in the isochrone, to systematic errors
in the $T_{eff}$ $vs.$ {\it (B--V)} calibration, or to a combination
of the two. To quantify the importance of this effect, let us suppose
that the assumed $T_{eff}$ $vs.$ {\it (B--V)} relation is correct and
that the mismatch is due solely to a systematic error in the theoretical
$T_{eff}$s of the giant stars, as given by the isochrones. In this case,
we can estimate by how much the $T_{eff}$s of the giants would need to be
increased (decreased) in order to shift the upper (lower) giant branch by
0.03 mag to the blue in Figure \ref{fig1}. With the aid of the $T_{eff}$
vs. {\it (B--V)} calibration derived in Paper I (Figure 4), we estimate
that the $T_{eff}$s of the giants need to be increased (decreased)
by $\sim$ 40 K in order to match the upper (lower) giant branch.

In the case of the Padova isochrones, the mismatch is different, the
theoretical isochrone being systematically $\sim$ 0.03--0.05 mag bluer
than the observations all along the giant branch. Thus, under the same
assumption as above, $T_{eff}$s predicted by the isochrones would need
to be systematically decreased by 40--80 K in order to match the observed
giant branch.

We note that such mismatches are not implausible in view of
discrepancies among current theoretical $T_{eff}$-scales predicted
by different groups, which amount to as much as 200 K (see Figure
2 of Schiavon, Barbuy \& Bruzual 2000, and discussion in Salaris et
al. 2002). The consequence of such a shift for our model predictions is
discussed in Section \ref{unc}.

\section{Isochrone-based Model Spectra vs. Observations: The
Spectroscopic Age of 47 Tucanae} \label{isoobs}

This section compares model line indices based on theoretical isochrones
to the observations. Here we briefly summarize the most important
ingredients of our models, referring the reader to Paper I for details.
Our models are based on a moderately-high resolution (1.8 {\AA} FWHM)
stellar spectral library (Jones 1999) and two sets of isochrones: those
by Salaris and collaborators, and the ones by Salasnich et al. (2000),
which are computed adopting the same chemical mixture as the former but
do not consider diffusion of heavy elements.  The observed integrated
spectra of 47 Tuc are described in Paper I.

\subsection{Models from the Original Isochrones} \label{modorig}

The procedure to compute integrated spectra from an observed CMD was
described in Paper I. The procedure used here is the same except that the
number of stars per isochrone point is given by theory plus a Salpeter
IMF, while in the computations from the CMD it is given by the observed
number of stars. Another difference refers to the $T_{eff}$s, which
are provided directly by the isochrones, so that they do not rely on a
{\it (B--V)} vs. $T_{eff}$ relation. However, we do need to resort to
such a relation when computing the weighted, fluxed spectrum for every
isochrone point.

Our first goal in computing isochrone-based models is to recalculate
the integrated model indices of Vazdekis et al. (2001) using the same
isochrones and spectral library, but adopting the new set of relations
derived in Paper I between fundamental stellar parameters ($T_{eff}$
and $L_{bol}$) and various observables (colors, absolute magnitudes,
spectra, and line indices).  Vazdekis et al. adopt the calibrations of
Alonso et al. (1995, 1996, 1999) for all evolutionary stages. We showed
in Paper I that our calibrations are consistent with Alonso et al.'s
for dwarfs but are roughly 70 K warmer at fixed {\it (B--V)} for giants.
Another difference is the stellar parameters of the library stars, which
were determined through different procedures. We will first concentrate
on the computations based on the Salaris isochrones, then contrast these
results with the computations from the Padova isochrones.

The model spectra are smoothed and rebinned to the resolution and
dispersion of the observed spectra of 47 Tuc, and the line indices
are measured using the definitions given in Paper I. Corrections
are applied as in Paper I, namely, corrections due to the paucity of
metal-poor giants in the library, the slightly lower metallicity of 47
Tuc (from {\it [Fe/H]} = --0.7 to --0.75), and the effect of CN-strong
stars. The isochrones adopted are computed for {\it [Fe/H]}=--0.7 and {\it
[$\alpha$/Fe]}=+0.4. However, in Paper I we inferred corrections to bring
line index predictions to {\it [Fe/H]}=--0.75, which is the effective
Fe abundance corresponding to our model predictions.  We also correct
the observed indices to eliminate the contribution by blue stragglers,
which of course are not included in the isochrones. The corrected EWs are
given in Table 4 of Paper I. Percentage corrections for blue stragglers
in each index are given in Table \ref{tbl-4}.

\begin{figure}
\plotone{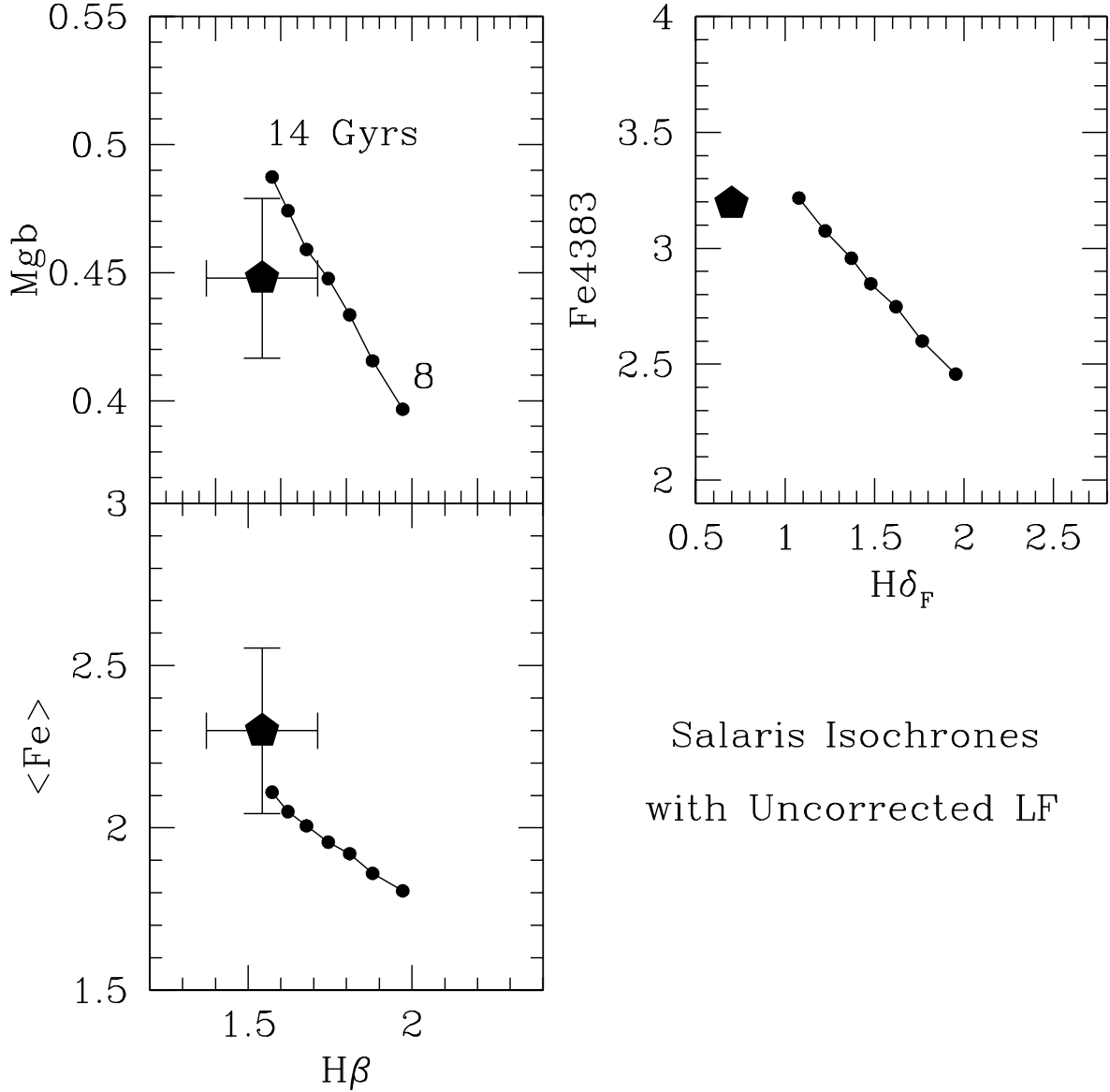}
\caption{Comparison between model predictions and observed absorption
line indices, corrected to eliminate the contribution by blue stragglers.
Small filled circles connected by solid lines represent the computations
based on the original Salaris isochrones, no AGB included, (see description in text)
for ages varying, from left to right, between 14 and 8 Gyr. The
isochrones have originally been computed for {\it [Fe/H]}=--0.7,
but model predictions are corrected, according to Paper I, to {\it
[Fe/H]}=--0.75. Other corrections to model predictions are also applied,
namely, for the paucity of metal-poor library giants and the
effect of CN-strong stars, also as described in Paper I. The best-fitting
spectroscopic age according to $H\beta$ is 14 Gyr (2-3 Gyrs older than the
best CMD-fitting age). For H$\delta_F$, the best fitting age is well in
excess of 14 Gyr. Finally, the observed indices have had the contribution
by blue stragglers subtracted, as indicated in Table 4 of Paper I.
\label{fig5} } 
\end{figure}

\begin{figure}
\plotone{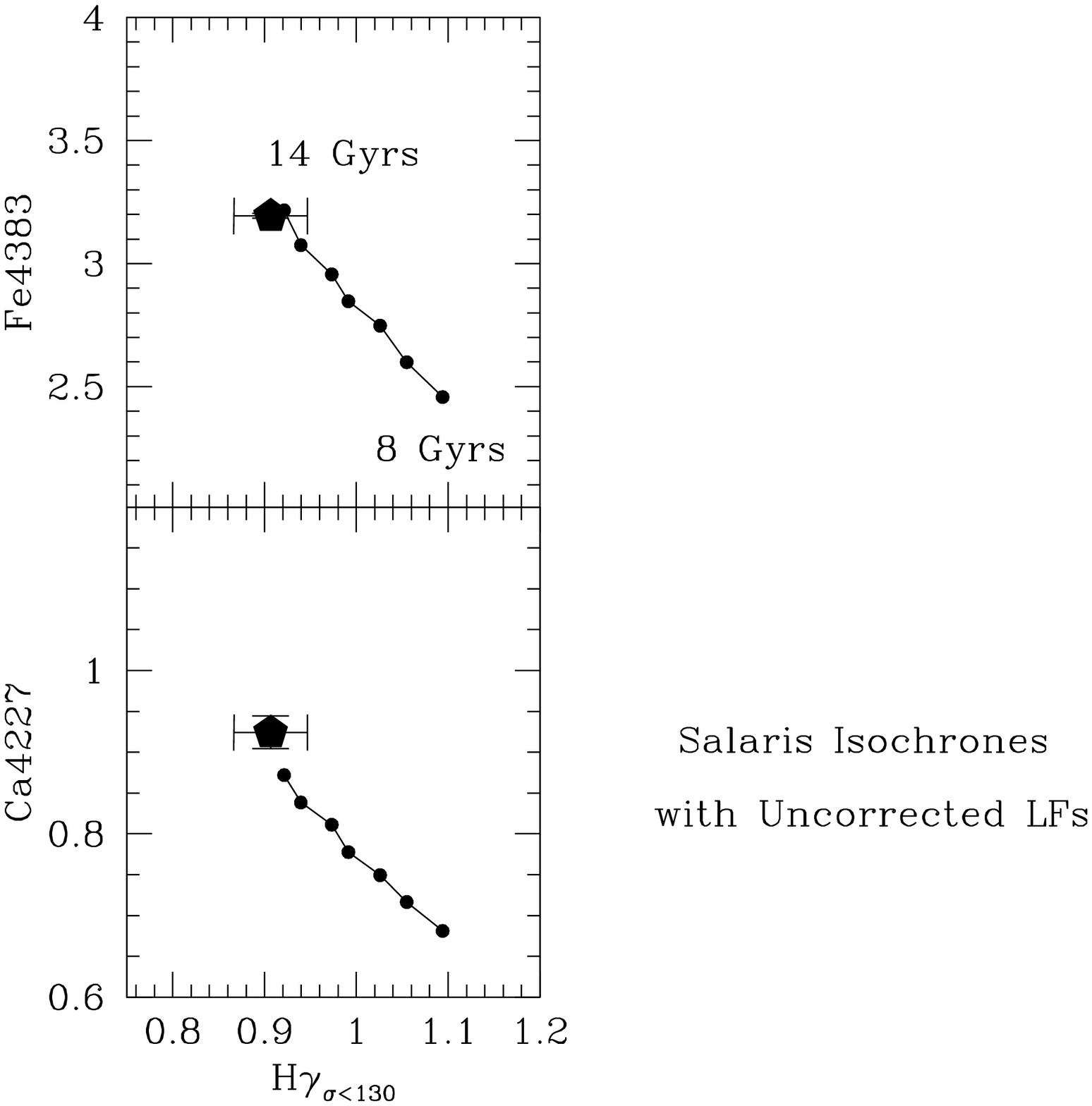}
\caption{Same as Figure \ref{fig5} for other combinations of
line indices. The models indicate an age of 14 Gyr
based on $H\gamma$.
\label{fig6}
}
\end{figure}

The first comparison is to Salaris models without AGB stars.  Figures
\ref{fig5} and \ref{fig6} show a comparison between these models versus
observed indices in a number of index $vs.$ index plots.  According to
these figures, different Balmer lines imply different ages. $H\beta$
and $H\gamma_{\sigma<130}$ are best fitted for an age of 14 Gyrs, while
$H\delta_F$ requires an even older age.  Most importantly, best-fitting
spectroscopic ages are always older (by at least $\sim$ 3 Gyrs) than that
obtained from the fit to the color and absolute magnitude of the cluster
turn-off (11--12 Gyrs). The models also provide good fits to all metal
lines, although the age for which the best fit is achieved again varies
from one metal line to the next. Note that the model prediction for {\it
(B--V)} is also too blue compared to the observations: for an age of 11
Gyrs, the model predicts {\it (B--V)}=0.79, which is 0.07 mag bluer than
the observed value (see Table 3 of Paper I). This again implies that an
older age is needed to reconcile with the observed color.  On balance,
these spectroscopic ages are consistent with the ones obtained by Vazdekis
et al. (2001) on the basis of the same set of isochrones and spectral
library but using a different set of stellar parameters and calibrations.

\begin{figure}
\plotone{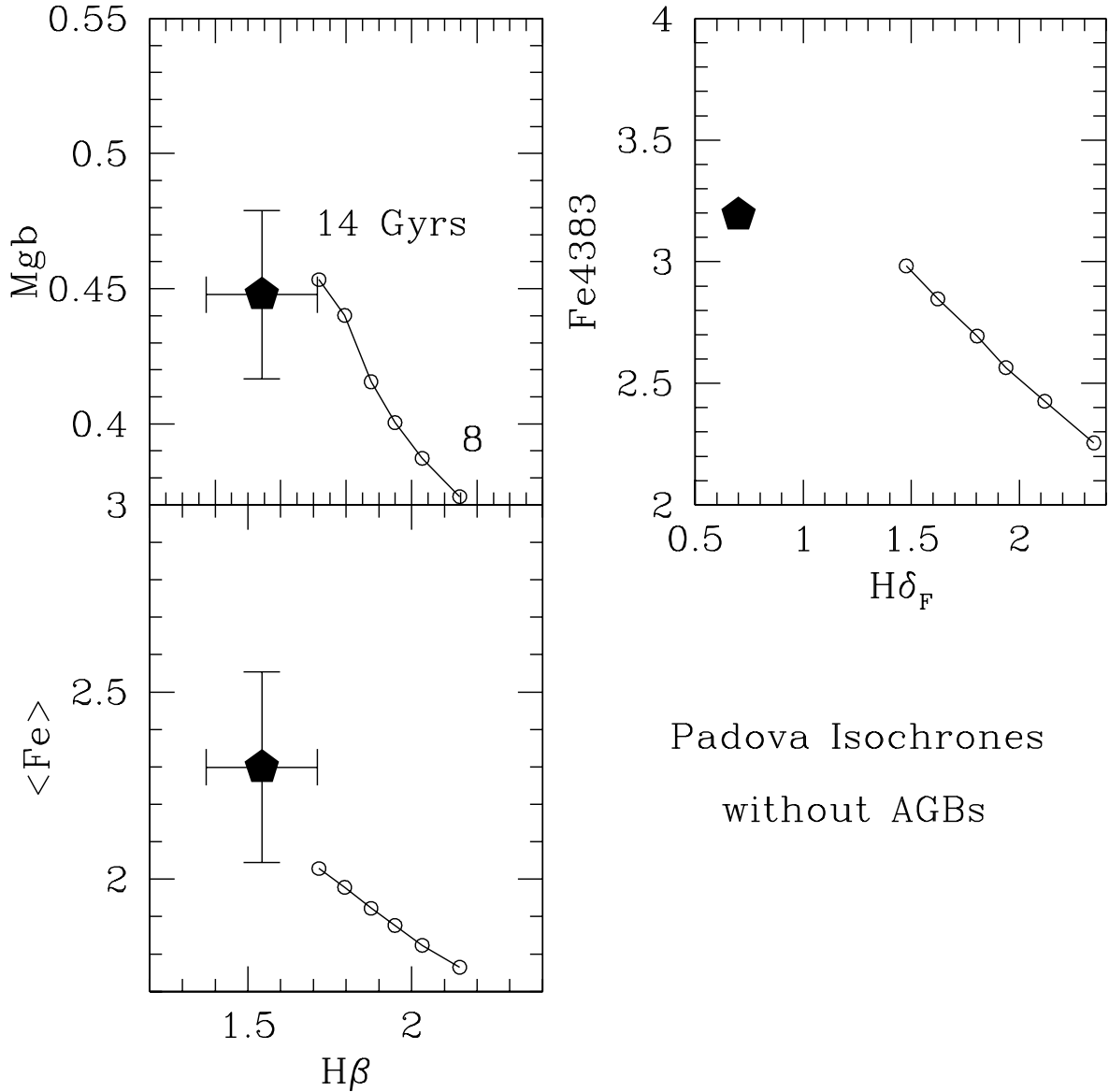}
\caption{Same as Figure \ref{fig5}, now showing model predictions from
the Salasnich et al. (2000) isochrones, excluding the contribution of
AGB stars. As expected, the ages inferred are older than those based on
Salaris isochrones by $\sim$ 2-3 Gyrs.  This is partly because Padova
isochrones do not consider diffusion of heavy elements, and partly
because they have slightly warmer giant branches (see text).
\label{fig7}
}
\end{figure}

\begin{figure}
\plotone{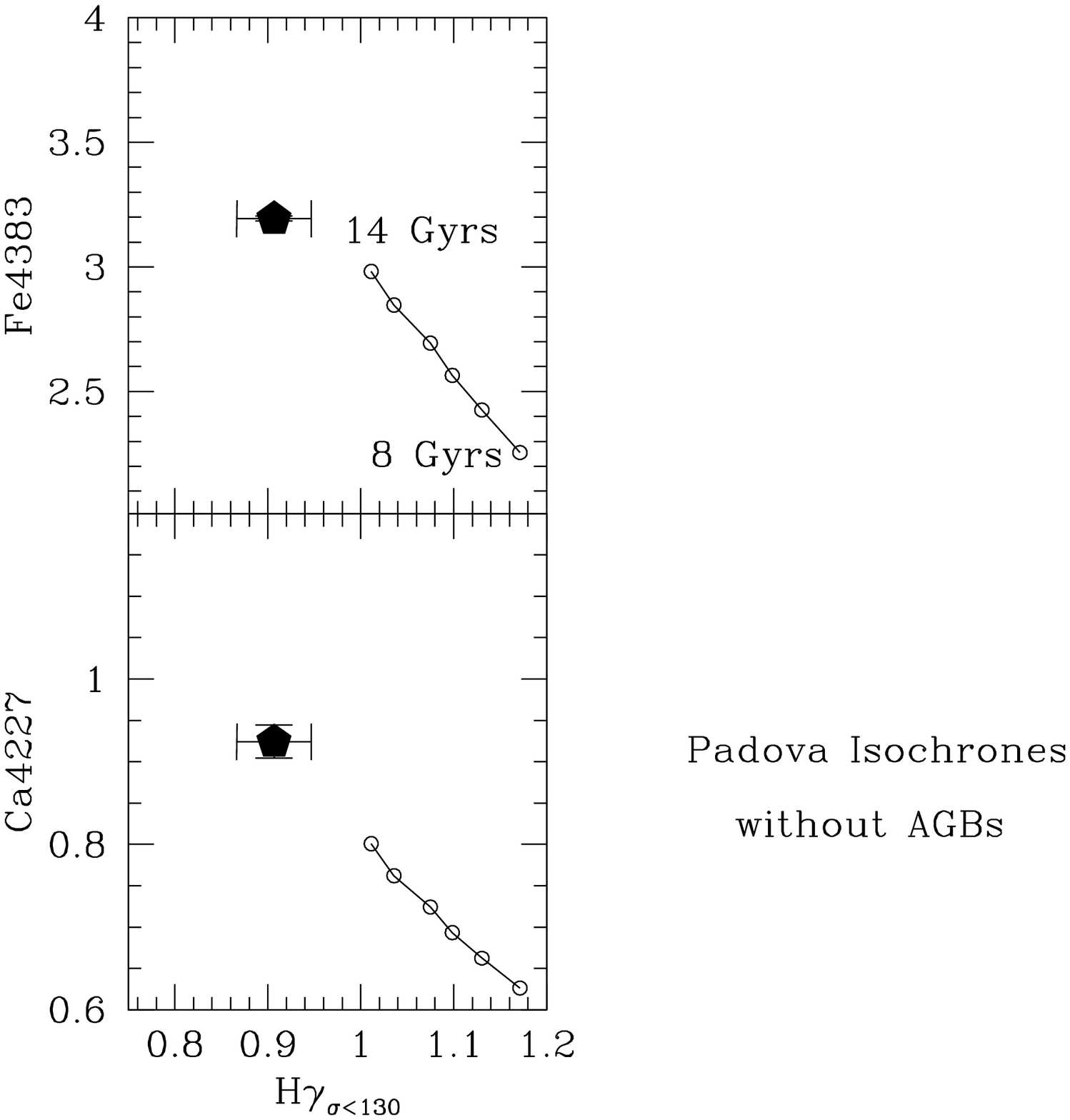}
\caption{Same as Figure \ref{fig7} for other combinations of
line indices. 
\label{fig8}
}
\end{figure}

Figures \ref{fig7} and \ref{fig8} reproduce the data of Figures
\ref{fig5} and \ref{fig6} but now using model predictions based on
the Padova isochrones. In order to bring all isochrones to the same
footing, we first compare results based on Padova isochrones after
removing their AGB components.  The basic conclusion from Figures
\ref{fig7} and \ref{fig8} is that spectroscopic ages inferred from
Padova isochrones are larger than those based on Salaris isochrones by
a little more than 2 Gyrs. Such a difference is significantly higher
than expected based on fits to the position of the cluster turn-off
(Figures \ref{fig1} and \ref{fig2}), which yielded an age difference
(due to heavy element diffusion) of $\sim$ 1 Gyr. The small discrepancy
is plausibly explained by the systematically too warm giant branch in
the Padova models, which forces an older overall spectroscopic age.
The temperature difference at the level of the Horizontal Branch is of
the order of 50 K and increases up to the tip of the Giant Branch, where
it reaches $\sim$ 100 K. This difference in the temperature of the red
giant branch is also responsible for another feature worthy of notice
in Figures \ref{fig12} and \ref{fig13}: that the metal lines are all
weaker when Padova isochrones are used (compare with Figures \ref{fig10}
and \ref{fig11}).

\subsection{Models with ``Corrected'' Luminosity Functions}

As discussed in Section \ref{isocmd}, both sets of isochrones employed
in this work underestimate the number of giant stars brighter than the
horizontal branch. In the case of the Salaris isochrones, which do not
take into account AGB stars, the mismatch is of the order of 0.4 dex,
while in the case of the Padova isochrones, which do include AGBs,
it is slightly lower, of the order of 0.2--0.3 dex. To evaluate the
impact of these underestimates on our computations, we apply empyrical
corrections to both theoretical LFs to bring them into agreement with
the observations.

\subsection{Salaris Isochrones and the Spectroscopic Age of 47 Tuc}

\begin{figure}
\plotone{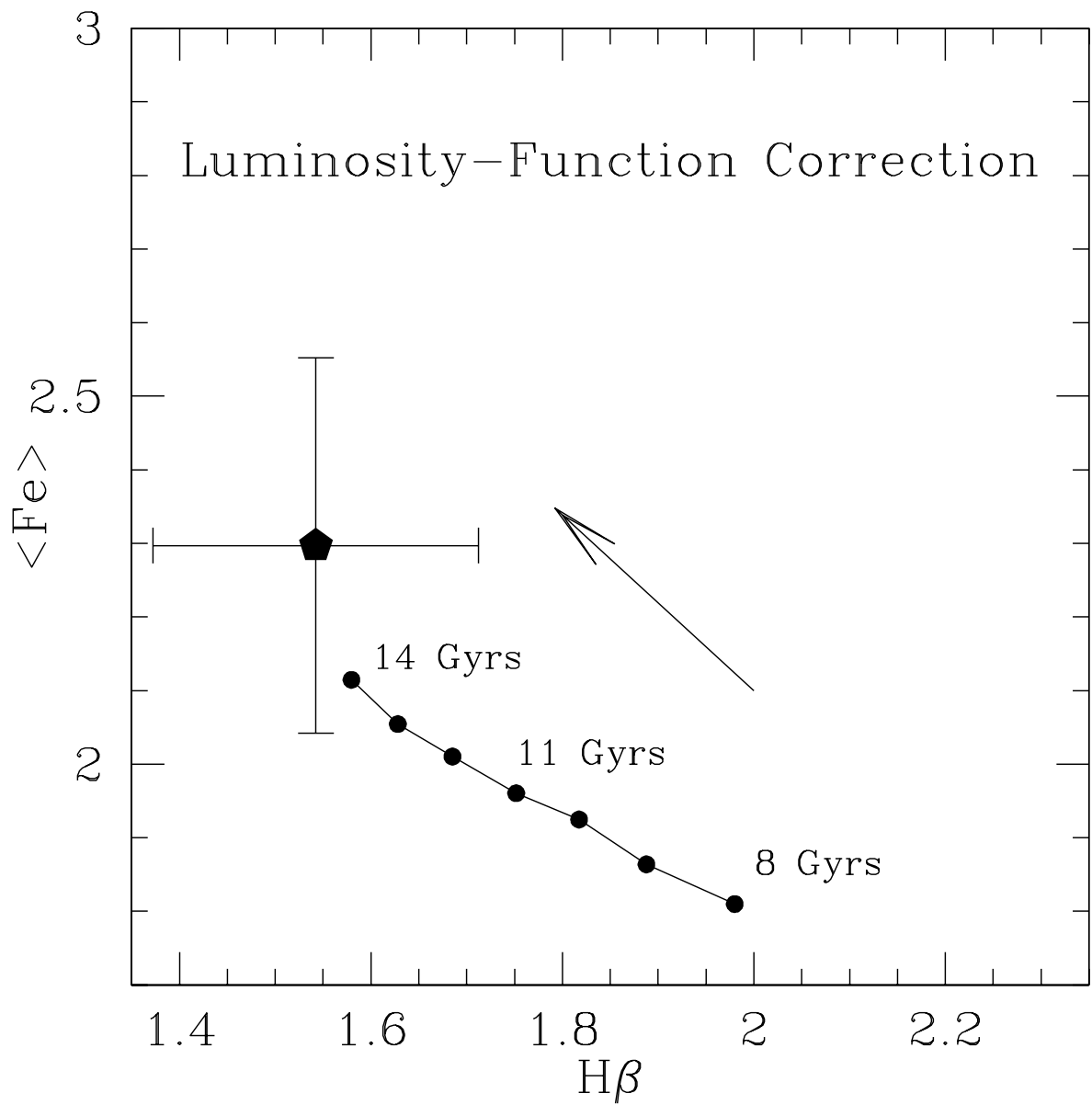}
\caption{The $<Fe>$ vs. $H\beta$ panel repeated from \ref{fig5}.
The solid dots are the model predictions using the Salaris isochrones
with their original, uncorrected LFs. The arrow indicates the extent
and direction by which the model predictions change if the theoretical
giant-branch LF is corrected according to the thin line in Figure
\ref{fig3}. In general, because of the increase of red-giant light
contribution, metal lines become stronger and hydrogen lines become
weaker. The new age estimated from $H\beta$ is now only 11-12 Gyrs,
bringing the spectroscopic age into good agreement with the isochrone
age from Figure \ref{fig1}.
\label{fig9}
}
\end{figure}

\begin{figure}
\plotone{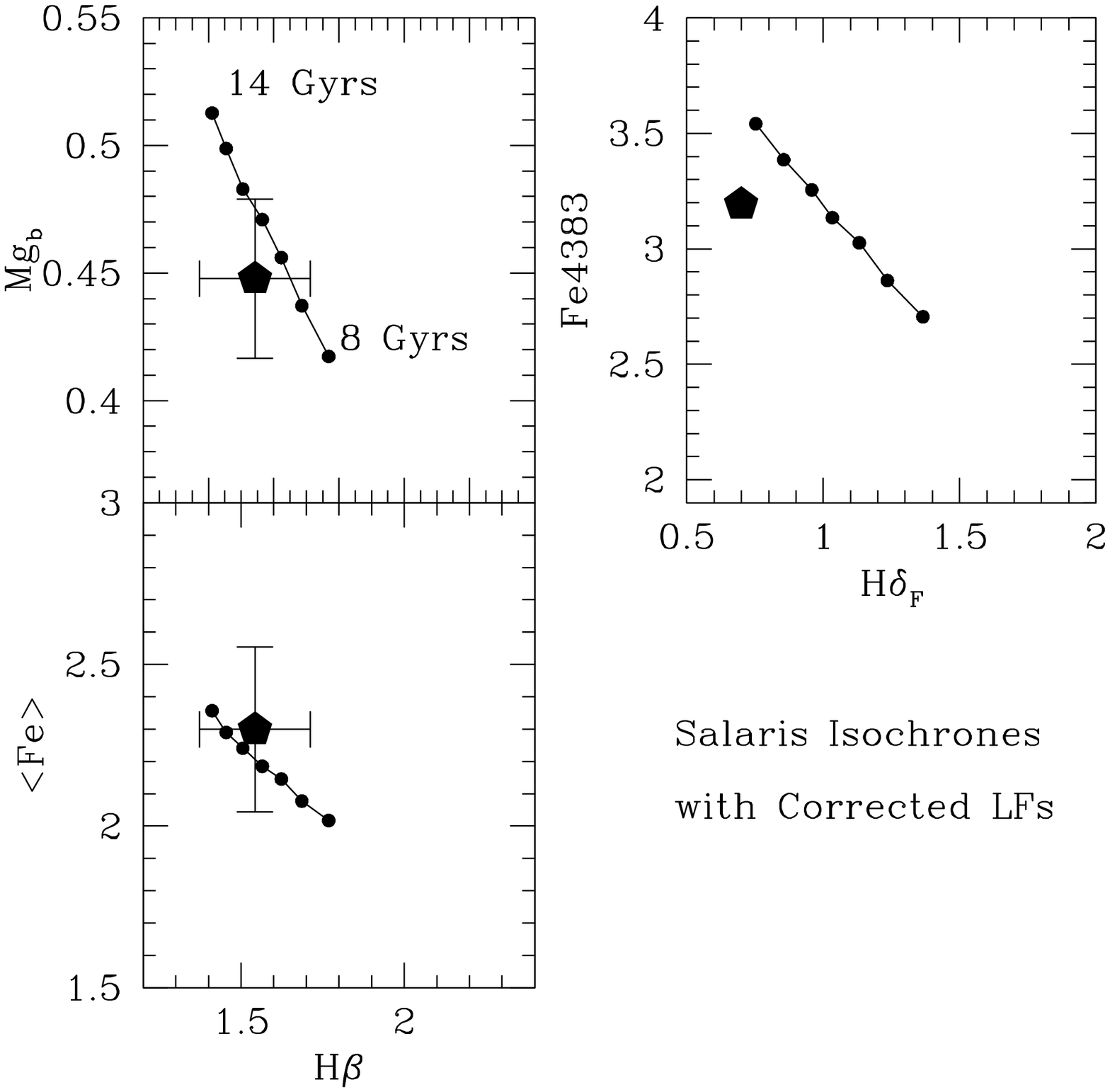}
\caption{Same as Figure \ref{fig5} but now using corrected LFs.  The
models shown in Figure \ref{fig5} have been shifted analogously to the
arrow in Figure \ref{fig9}.  The correction brings the spectroscopic age
for $H\beta$ into agreement with CMD-based age $\sim$ 11 Gyrs. H$\delta_F$
still gives a slightly higher age ($\sim$ 14 Gyrs), as discussed in the
text and in Paper I.
\label{fig10}
}
\end{figure}

\begin{figure}
\plotone{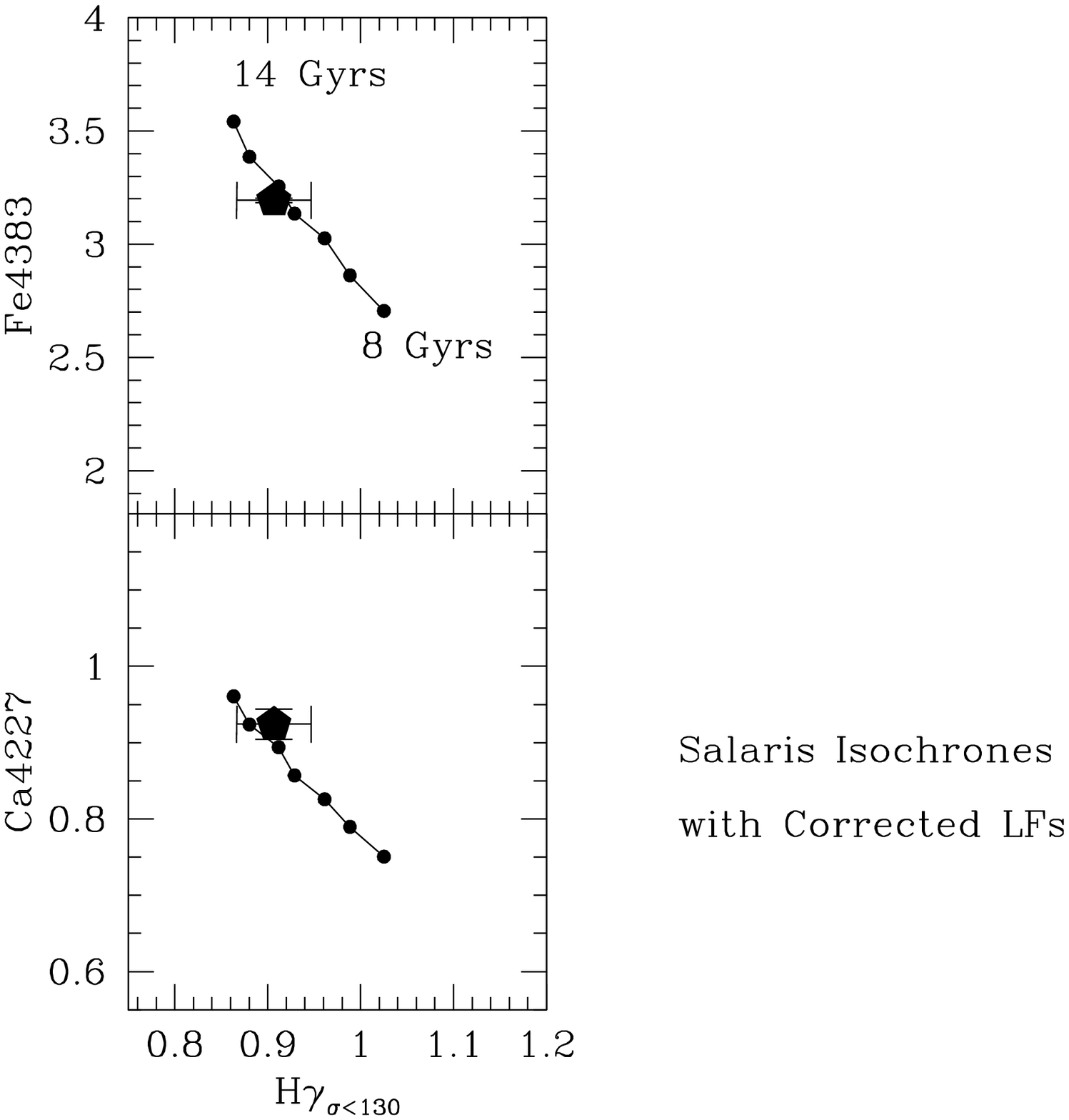}
\caption{Same as Figure \ref{fig10}, for other combinations of line
indices. For $H\gamma_{\sigma<130}$ the models predict an age in agreement
with the CMD-based and $H\beta$-based ages, namely, 12 Gyrs.
\label{fig11}
}
\end{figure}

\begin{figure}
\plotone{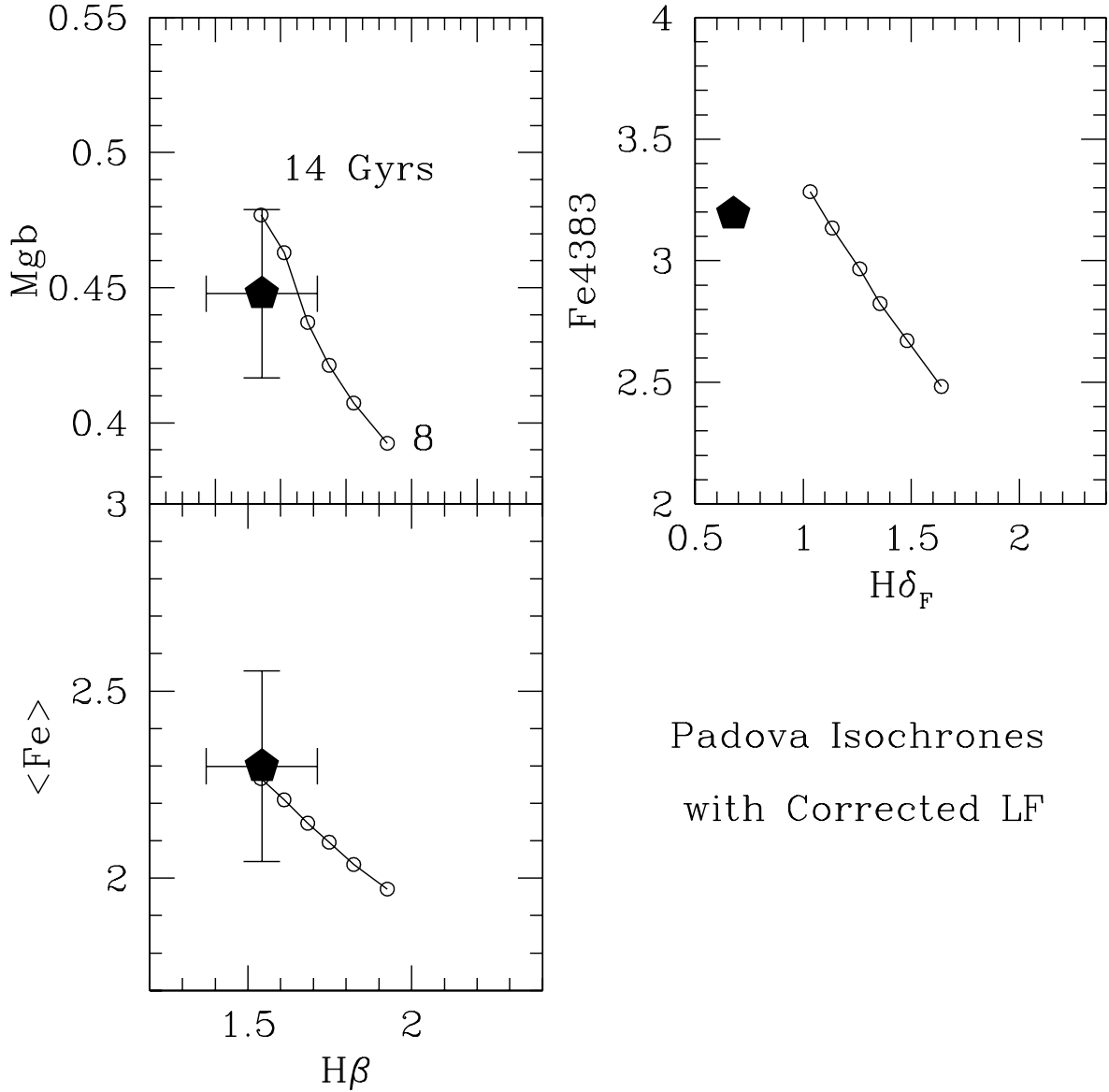}
\caption{Same as Figure \ref{fig7} but now using corrected LFs. The
models shown in Figure \ref{fig7} have been shifted analogously to the
arrow in Figure \ref{fig9}. The correction reduces the spectroscopic age
for $H\beta$ to $\sim$ 14 Gyrs, in agreement with the CMD-based age for
Padova isochrones (\ref{fig2}). As in the case of Salaris isochrones, 
the spectroscopic age based on H$\delta_F$ is higher.
\label{fig12}
}
\end{figure}

\begin{figure}
\plotone{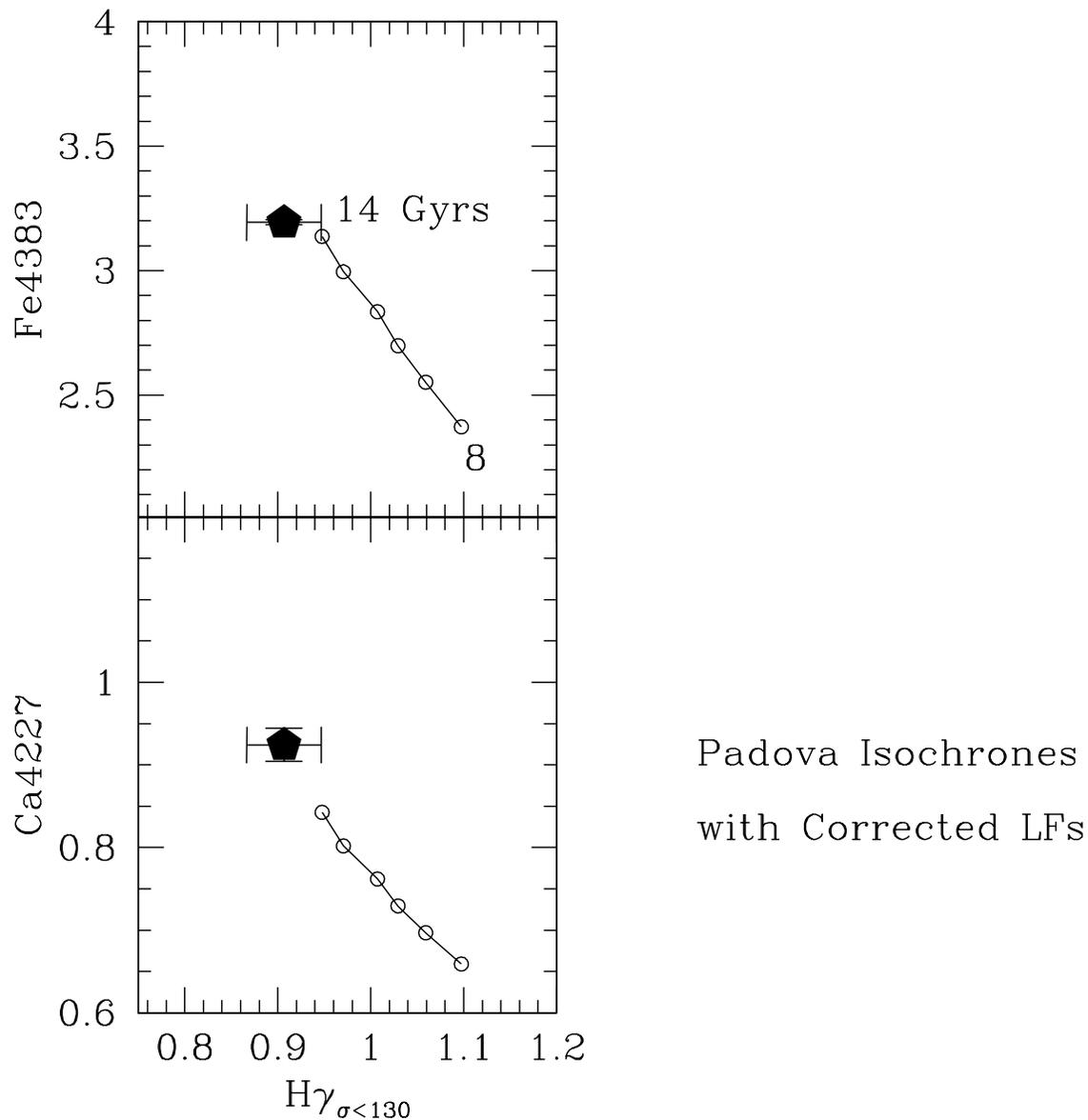}
\caption{Same as Figure \ref{fig12}, for other combinations of line
indices. For $H\gamma_{\sigma<130}$ the LF-corrected Padova models
predict a spectroscopic age which is consistent with the CMD-based and
$H\beta$-based ages, namely, $\sim$ 14 Gyrs. Because of the absence of
heavy-element diffusion in the Padova isochrones, and because their giant
branches are slightly warmer, the spectroscopic age inferred is older
than the one based on Salaris isochrones by $\sim$ 2 Gyrs.
\label{fig13}
}
\end{figure}

We first correct the Salaris isochrones in order to bring their
predicted LF into agreement with the observations; discussion of the
Padova isochrones is postponed to the next sub-section.  The corrected
Salaris LF for a model with 11 Gyrs is represented in Figure \ref{fig3}
as the thin line.  The consequence of correcting our model predictions
for this effect is illustrated as the arrow in Figure \ref{fig9} for the
$<Fe>$--$H\beta$ diagram. Since the giant light contribution is increased,
the integrated spectrum becomes redder, metal lines become stronger,
and hydrogen lines become weaker. With this correction to the LF, the
spectroscopic age based on $H\beta$ is reduced from $\simgreat$ 14 to 11
Gyr, in excellent agreement with the isochrone age in Figure \ref{fig1}.

We checked the size of the corrections by recomputing the integrated
spectrum directly from the CMD as in Paper I, but excluding AGB/RGB stars
so as to force agreement with the theoretical uncorrected luminosity
function. Differences between the line indices obtained in Paper I and
the ones computed with the exclusion of AGB/RGB stars are essentially the
same as the ones obtained adopting the two LFs shown in Figure \ref{fig3},
which validates the corrections estimated here.

Figures \ref{fig10} and \ref{fig11} compare the LF-corrected model
predictions with the observations for the other indices.  Virtually all
line indices are well fit for ages ranging from 9 to 14 Gyrs.  The best
fitting age for $H\gamma_{\sigma<130}$ is 11--12 Gyrs, in agreement
with the age inferred from $H\beta$ and CMD-fitting. The notable
exception is $H\delta_F$, for which the best-fitting age is slightly
higher: $\sim$ 14 Gyrs. The latter result was anticipated in Paper I,
because the model spectrum computed directly from the CMD of the cluster
also overestimates $H\delta_F$, and when we considered the correction
of $H\delta_F$ due to the effect of CN-strong stars, the mismatch was
further increased.  However, as discussed in Section 5.3 of Paper I,
we have reason to believe that our correction for this effect on
$H\delta_F$ may be in error. In any case it was already clear in Paper I that
$H\delta_F$ was aberrant. All the metal-line indices are well fit by
models with ages of 11-12 Gyrs, including Ca4227, which was greatly
improved after correction for the effect of CN-line contamination,
derived in Paper I. Finally, the integrated {\it (B--V)} of the cluster
is also matched for an age of 11 Gyrs.

The ratio of the contribution to the integrated light of turn-off stars
to giant stars (HB stars excluded) provides a good gauge of the impact
of the underestimate of the number of bright stars in the theoretical
isochrone. In Paper I we computed the fractional contribution to the
integrated light by different evolutionary stages at a number of reference
wavelengths. Those numbers are given in Table 5 of Paper I.  They are
fairly robust, since they are based on a statistically representative
CMD of the cluster. From Table 5 of Paper I, it can be seen that the
ratio of turn-off light to giant light is 0.50 and 0.28 in the regions
of $H\delta_F$ and $H\beta$ respectively. If the uncorrected theoretical
LF is adopted, those figures change to 0.77 and 0.47 respectively.

We have just shown that a 0.4 dex underestimate in the contribution of
red giants to the integrated light leads to an age overestimate of about
3--4 Gyrs for fixed hydrogen line equivalent width. It is interesting to
note that this age difference is the same as the one found by Vazdekis et
al. (2001) between their spectroscopic and CMD-based age determinations.

The fact that the amount of red giant light contribution has such an
important effect on spectroscopic age estimates may be surprising,
given that it is often said in the literature that spectral features
in the integrated light of old stellar populations evolve with age
mainly as a function of the temperature of the turn-off stars. Though
this statement is strictly correct, it neglects the basic fact that a
significant fraction, if not most, of the {\it continuum} light of old
stellar populations even in the blue is provided by giant stars (see
the discussion in Section 5.5 of Paper I). Thus, our results show that
the luminosity function is a key constraint on population synthesis
models, upon which our ability to correctly predict the ages of old
stellar populations from their integrated light is strongly dependent.

In summary, we conclude that the too-large spectroscopic age of 47 Tuc
found in previous works is due to a combination of reasons: the neglect
of important effects like $\alpha$-enhancement and He-diffusion (as
previously shown by Vazdekis et al. 2001), and the underestimate of the
contribution of bright red giants to the integrated light of the cluster.

\subsection{Padova Isochrones} \label{padova}

In this section we compare the observed absorption line EWs with model
predictions based on Padova isochrones, now including AGB stars. As
in the previous section, the LF above the Horizontal Branch has been
empyrically corrected in order to reproduce the observed number
counts. The comparisons are displayed in Figures \ref{fig12} and
\ref{fig13}, where it can be seen that the spectroscopic age inferred
when the LF-corrected Padova isochrones are adopted is about 14 Gyrs,
for both $H\beta$ and $H\gamma$ ($H\delta_F$ again being older).  Based on
Section \ref{modorig}, we expect that the Padova isochrones will yield a
spectroscopic age that is roughly 1--2 Gyrs older than the one that is
inferred when using the Salaris isochrones. 

Aside from these systematically older ages, it is important to emphasize
that, for $H\beta$ and $H\gamma$ we achieve generally good consistency
between spectroscopic and CMD-based ages for the Padova models, just as we
did with the Salaris models. As for the model predictions for $H\delta_F$
they are again much higher than the observations, thus leading to too
old ages according to this index.

We also performed computations using the original Padova isochrones, i.e.,
including AGB stars with their theoretically predicted LF (solid line in
Figure \ref{fig4}), which underpredicts the observed LF of bright giants
by 0.2--0.3 dex. The spectroscopic age inferred in this case is $\sim$
2--3 Gyrs older than the one inferred above, and thus is in disagreement
with the CMD-based age. Therefore, we again stress the importance of
correctly matching the giant branch luminosity function in order to
predict spectroscopic ages which agree with the ones based on the fit
of the cluster CMD. As discussed in previous sections, it is hard to
quantify to which extent the LF mismatch is due to errors in the models
predictions of AGB and first-ascent RGB stars. While AGB stars may have
larger uncertainties, we have reasons to believe that the predictions for
first-ascent giants may be also in error, as noted in Section \ref{lfun}.

\section{Uncertainties and Systematic Effects due to Input Stellar
Parameters} \label{unc}

In this section we gauge the effects of uncertainties in the $T_{eff}$ and
{\it [Fe/H]}-scales adopted as model inputs. As discussed in Paper I,
the $T_{eff}$ and {\it [Fe/H]}-scales of metal-poor stars from different
sources in the literature may differ by up to $\sim$100K and 0.1 dex.
We also argued in Section ~\ref{isocmd} that the $T_{eff}$s predicted by
theoretical isochrones may be in error by comparable amounts.  Our set
of stellar parameters was determined from Str\"omgren photometry for the
dwarfs and from calibrations of line-indices $vs.$ stellar parameters
for the giants. Our $T_{eff}$ and metallicity-scales for the library
giants are, by construction, tied to the stellar parameter calibrations
of Soubiran et al. (1998) (see Paper I for details).

As in Paper I, we resort to fitting functions in order to assess
the changes in line indices that result from slight changes in model
inputs.  The stellar parameters tested were the $T_{eff}$ and {\it
[Fe/H]}-scales of dwarfs and giants separately, plus the assumed {\it
[Fe/H]} of 47 Tuc. For these computations, giants are considered to be
all stars with $T_{eff} <$ 6000 K and $\log g <$ 3.6, and dwarfs are
in the same $T_{eff}$ interval but with $\log g >$ 3.6 (thus including
subgiants). From the discussion of systematic effects on these stellar
parameters carried out in Paper I, we chose to vary the $T_{eff}$s of
giants and dwarfs by $\pm$ 75 and $\pm$ 50 K respectively. The {\it
[Fe/H]}s of giants and dwarfs were varied by $\pm$ 0.1 and $\pm$ 0.05
dex respectively, while the {\it [Fe/H]} of 47 Tuc was varied by $\pm$
0.1 dex.

The results are displayed in Table \ref{tbl-4}, where we give the
percentage variation of each line index as a function of model input
changes. To aid the reader in judging the importance of these effects,
we provide in the last row of Table \ref{tbl-4} the percentage variations
due to an increment of 1 Gyr in age (from 10 to 11 Gyrs).  The best way to
interpret the numbers in Table \ref{tbl-4} is to compare the percentage
variation in each index with the variation due to the age change given
in the last row.  For instance, $H\delta_F$ varies by $\pm$ 9\% for an
uncertainty of $\pm$ 50 K in the $T_{eff}$s of dwarfs. It varies by --7\%
when age changes from 10 to 11 Gyrs. Therefore, the $\pm$ 50 K uncertainty
in the $T_{eff}$s of dwarfs implies a $\mp$ 1.3 Gyr uncertainty in the
age inferred from $H\delta_F$ at the metallicity of 47 Tuc.

%This is a very important effect, in view
%of the fact that the $T_{eff}$ of the model giant branches can differ by
%as much as 200 K between different sets of isochrones (see Salaris 2001,
%and Schiavon, Barbuy \& Bruzual 2000).

Before discussing systematic effects in the input stellar parameters,
we assess the effect of the mismatch of the theoretical isochrones to
the red giant branch of 47 Tuc in Figures \ref{fig1} and \ref{fig2}. In
Section \ref{cmd} we saw that the Salaris isochrone is bluer by $\sim$
0.04 mag at the lower giant branch and redder by a comparable amount
at the upper giant branch. In the case of the Padova isochrones,
the theoretical prediction is bluer than the observations through all
the RGB, by $\sim$ 0.03 mag. Such color differences, if due solely to
temperature effects, translate into $T_{eff}$ mismatches of $\pm$ 50
K in the case of the Salaris isochrones and + 70 K in the case of the
Padova isochrones, according to the calibration derived in Paper I. In
the case of the Salaris isochrones, correcting for these mismatches
amounts to negligible changes in the line index predictions. In fact,
the percentage variations in all indices correspond to less than 0.4 Gyr,
so that we do not need to perform any correction to our model predictions
due to the slight mismatch in the red giant branch. This is because
there is a compensation between the opposite $T_{eff}$ mismatches in
the lower and upper giant branch. In the case of the Padova isochrones,
as the model RGB is systematically bluer than the observations, the
effect on spectroscopic age predictions is not negligible, leading to
systematic age overestimates of about 1 Gyr. In section \ref{padova}
we inferred a spectroscopic age of 14 Gyrs from $H\beta$ and slightly
above that value from $H\gamma$. Correcting these predictions for the
systematic temperature error in the RGB would bring them into better
agreement with the age inferred from the position of the turn-off found
in Section \ref{cmd} (12.5 $<$ t $<$ 14.1 Gyrs).

Uncertainties due to systematic errors in the input parameters of
dwarf stars are smaller than the variations of the input parameters
of giants. Thus, they have only a relatively minor impact on model
predictions, at the old ages of GCs. Except for $H\delta_F$, all 
line indices vary by less than 3\% when varying the $T_{eff}$ and {\it
[Fe/H]}-scales of dwarfs. Taking $H\beta$ as illustrative, this translates
to roughly a 0.5 Gyr uncertainty in age due to the $T_{eff}$-scale
of dwarfs and a 0.25 Gyr uncertainty due to their {\it [Fe/H]}-scale.
By comparison, uncertainties of 75 K or 0.1 dex in the input $T_{eff}$
and {\it [Fe/H]}-scales of giants produce a 1 Gyr uncertainty in the
age inferred from either $H\beta$ or $H\delta_F$. This is particularly
important in view of the current discrepancies in the $T_{eff}$-scales
of giant stars among different sets of the most recent theoretical
isochrones, which amount to 200 K (Salaris et al. 2002, Schiavon,
Barbuy \& Bruzual 2000). Such a large discrepancy is still lingering,
mostly because the position of the theoretical red giant branch is
dependent on poorly constrained inputs of stellar evolutionary models,
such as the mixing length parameter and the T($\tau$) relation adopted
as boundary condition for the interior models. On the other hand, the
present work may be viewed as an important {\it consistency check} on
the current theoretical Salaris $T_{eff}$s, as Table \ref{tbl-4} shows
that an error of $\pm$ 200 K in the $T_{eff}$s of the giants would cause
a mismatch of $\mp$3 Gyr between the CMD and the spectroscopic ages,
which is not seen. This is a valuable independent confirmation of the
correctness of the Salaris $T_{eff}$-scale, at least for giants.

The input metallicity of 47 Tuc is also an important source of uncertainty
for spectroscopic age determination. Varying {\it [Fe/H]}$_{47\,Tuc}$
by $\pm$0.1 dex (the nominal errorbar quoted by Carretta \& Gratton
1997) causes the spectroscopic age to vary by roughly $\pm$ 1 Gyr, as
measured from all Balmer lines.  The effect of blue stragglers is also not
negligible. If the observations are not corrected for the contribution
of blue stragglers, ages are underestimated by slightly more than 1 Gyr
when measured from $H\delta_F$.  The effect on $H\gamma_{\sigma<130}$
and $H\beta$ is less important, leading to underestimates of $\sim$
0.6 and 0.5 Gyrs respectively.

\section{Conclusions} \label{conc}

In this paper, we compared model single stellar population spectra to
the integrated spectrum of the metal-rich Galactic globular cluster 47
Tuc. The model spectra were computed from two sets of state-of-the-art
theoretical isochrones.  We tested the sensitivity of our predictions
to variations in a number of key model inputs related to the spectral
library employed, the theoretical isochrones, and the relations between
fundamental and observed stellar parameters. Our main conclusions can
be summarized as follows:

-- Different sets of isochrones predict slightly different ages when
compared to the cluster data on the color-magnitude diagram.  In the
case of the Salaris set, the isochrone providing the best match to the
turn-off of the cluster has an age of 11-12 Gyrs. Vazdekis et al. (2001),
using the same set of isochrones, obtained an age younger by about 1
Gyr. The difference is minor and is due to the different reddenings
assumed. When the isochrones of Salasnich et al. (2000) are used, a
slightly older age is obtained (12.5-14 Gyrs) the reason being that the
latter set of isochrones does not consider diffusion of heavy elements.
A final decision as to which set of isochrones is best awaits a better
knowledge of the extent to which atomic diffusion is operating in main
sequence stars.

-- When compared to the observations in the luminosity function domain,
both sets of theoretical isochrones adopted in this work underestimate
the number of giant stars brighter than the horizontal branch. In the
case of the Salaris isochrones, which do not include AGB stars, the
mismatch amounts to 0.3--0.4 dex. The Padova set, even though it does
include AGB stars, is also offset, though by a smaller amount ($\sim$
0.2--0.3 dex). It is unclear to us what causes this discrepancy.  At the
level of the AGB bump, roughly 2/3 of the mismatch seems to be due to the
lack of AGB stars in the isochrones, this fraction being probably lower
in other parts of the upper giant branch. However, even after performing
an approximate correction for the absence of AGB stars, there appears
to be a small residual discrepancy between models and observations
for first-ascent giants. We conclude that theoretical predictions to
the luminosity functions of both the RGB and the AGB may be too low,
though it is difficult to quantify by how much in each case.

-- Spectroscopic ages inferred from $H\beta$ and $H\gamma$ when
adopting isochrones without any AGB stars are systematically older, by
at least 3 Gyrs, than the ages inferred from the fit to the cluster's
CMD. Moreover, metal-lines are systematically too weak and the
integrated {\it (B--V)} is too blue by 0.07 mag.

-- Empirically correcting the theoretical LFs to match the observed
luminosity function reduces the spectroscopic age to 11--12 Gyrs 
in the case of the Salaris isochrones, and $\sim$ 13 Gyrs in the case of
the Padova ones as determined from $H\beta$ and $H\gamma$ (the latter
is obtained after correction for a systematic mismatch of the cluster's
RGB by Padova isochrones). Such revised ages are now in excellent
agreement with the ages inferred from fitting both sets of isochrones
to the CMD.  Also {\it (B--V)} and metal-lines studied are very well
fit by the LF-corrected models.

-- The best value found from the fit to $H\delta_F$ is still higher by
$\sim$ 3 Gyrs than found for the other Balmer lines.  This mismatch is
in the same sense found in Paper I, where $H\delta_F$ computed directly
from the CMD-based synthesis was relatively stronger than $H\gamma$
or $H\beta$.  As $H\delta_F$ is more strongly affected by C,N abundance
anomalies, we suggested in Paper I that the discrepancy was due to our
failure to properly model and correct for CN lines within the passband
and continuum windows of the index.

-- We estimate the impact on spectroscopic ages of uncertainties in a
number of model inputs, such as the $T_{eff}$ and {\it [Fe/H]}-scales
of dwarfs and giants, the luminosity function of the red giant branch
and the metallicity of the cluster.  Spectroscopic age-dating of old
stellar populations is mostly influenced by uncertainties coming from
the $T_{eff}$ and {\it [Fe/H]}-scale of giant stars, as well as the
luminosity function of the upper red giant branch and the metallicity
adopted for the cluster.  Using $H\beta$, a 1 Gyr uncertainty is caused
by an error of 75 K in the giant $T_{eff}$-scale, a 0.1 dex error in
the giant-branch luminosity function, and a 0.1 dex error in either
the {\it [Fe/H]}-scale of giants or the {\it [Fe/H]} of the stellar
population. Because the uncertainties in the parameters of dwarf stars
are lower, their impact on model predictions is generally less important.

-- The contribution by blue stragglers to the integrated line
indices is important only for $H\delta_F$ and slightly less so
for $H\gamma_{\sigma<130}$. When $H\delta_F$ is corrected for the
contribution of blue stragglers, we obtain an age 1 Gyr older. In the
case of $H\gamma_{\sigma<130}$, the age inferred is 0.5 Gyr older.

In summary, we conclude that the exceedingly high spectroscopic ages
found for 47 Tuc in previous works was due to: 1) the use of theoretical
isochrones that neglect important effects such as He-diffusion and
$\alpha$-enhancement, as shown by Vazdekis et al. (2001), and 2) the
underestimate of the number of giant stars above the horizontal branch,
which was partly due to the omission of AGB stars from the isochrones,
and partly to a genuine apparent shortfall in the predicted numbers
of AGB stars and, most likely, first-ascent giants as well.  Another
requirement, which is discussed in Paper I, is on the iron abundance of
the cluster. Models provide a good match if {\it [Fe/H]} is 0.05 dex
lower than found by Carretta \& Gratton (1997). Such a shift in {\it
[Fe/H]} is well within the errorbars of Carretta \& Gratton's analysis
and the uncertainties in the {\it [Fe/H]}-scale of our stellar library.

With regard to our long term goal of constructing metal-rich stellar
population models for elliptical galaxies, we have made considerable
progress.  First, the ages from $H\beta$ and $H\gamma_{\sigma<130}$ agree,
after the latter is corrected for the effect of C,N abundance variations
(see Paper I). Second, all metal lines are correctly predicted for the
same ages obtained from Balmer lines. Third, the giant temperature scale
of the Salaris models looks to be generally correct. On the other hand,
$H\delta_F$ is shown to be a problematic feature, embedded as it is in a
forest of CN lines. This problem is likely to be further exacerbated in
elliptical galaxies, which have notoriously strong CN features. Moreover,
the present models do not include variations in Type Ia vs. Type II
supernovae element ratios (Trager et al. 1998), nor do they probe
the high-metallicity regime needed for ellipticals. Finally, we have
identified an error in the upper giant branch luminosity function of
current evolutionary models, which underestimate the aggregate number
of luminous red giants in 47 Tuc by roughly a factor of 2. We have
furthermore shown that an accurate LF for these stars is needed for
accurate spectroscopic age determination. A major unknown is whether the
red-giant excess of 47 Tuc extends to all metal-rich old populations,
and what its magnitude might be. Testing the luminosity function in old
metal-rich populations has emerged as an important prerequisite for
stellar population studies.  These and further improvements are left
for future works.

\acknowledgments
We would like to thank Maurizio Salaris for making his isochrones
available.  Stefano Covino is thanked for the red integrated spectrum of
47 Tuc. We would also like to thank Alexandre Vazdekis, Maurizio Salaris,
Manuela Zoccali, Achim Weiss and Mike Bolte for helpful discussions. The
referee, Brad Gibson, is thanked for valuable suggestions that greatly
improved this paper. R.P.S. thanks the hospitality of the Physics
Dept. of the University of North Carolina, Chapel Hill, where part
of this work was developed. Likewise, J.A.R. thanks the Astronomy
Department at UC, Santa Cruz for hospitality during a visit in which
part of this work was developed. This work has made extensive use
of the Simbad database. R.P.S. acknowledges support provided by the
National Science Foundation through grant GF-1002-99 and from the
Association of Universities for Research in Astronomy, Inc., under NSF
cooperative agreement AST 96-13615, and CNPq/Brazil, for financial help
(200510/99-1). This research has also been supported by NSF grant
AST-9900720 to the University of North Carolina, and by NSF grants
AST-9529098 and AST-0071198 to the University of California, Santa Cruz.

\clearpage

\clearpage

\clearpage

\clearpage

\clearpage

\clearpage

\clearpage

\clearpage

\clearpage

\clearpage

\clearpage

\clearpage

\clearpage

\begin{deluxetable}{ccccccccc}
\tablecaption{Variation of line indices (\%) as a function of uncertainties
in model input parameters. In the last row, percentage variations are
given for a change from 10 to 11 Gyrs, according to our models. 
\label{tbl-4}}
\tablewidth{0pt}
\tablehead{
\colhead {} $H\delta_F$ & Ca4227 & $H\gamma_{\sigma<130}$ &
Fe4383 & $H\beta$ & $Mg_b$ & $Mg_2$ & $<Fe>$ &  }
\startdata

$\pm$ 4 & $\mp$ 11 & $\pm$ 1.5 & $\mp$ 4 & $\pm$ 4.5 & $\mp$ 6 &
$\mp$ 8.5 & $\mp$ 3.5 & $T_{Giants} \pm$ 75 K\\

$\pm$ 9 & $\mp$ 1 & $\pm$ 1.5 & $\mp$ 1.5 & $\pm$ 1.5 & $\mp$ 1 &
$\mp$ 1 & $\mp$ $<$ 1 & $T_{Dwarfs} \pm$ 50 K\\

$\mp$ 11 & $\pm$ 2 & $\pm$ 2 & $\pm$ 5 & $\pm$ 3 & $\pm$ 3.5 & 
$\pm$ 5.5 & $\pm$ 4.5 & $[Fe/H]_{Giants} \pm$ 0.1\\

$\mp$ 2.5 & $\pm$ 1 & $\pm$ 1 & $\pm$ 1.5 & $\pm$ $<$ 1 & $\pm$ 1 & 
$\pm$ 1 & $\pm$ 1 & $[Fe/H]_{Dwarfs} \pm$ 0.05\\

$\mp$ 7.5 & $\pm$ 2.5 & $\pm$ 4 & $\pm$ 6 & $\pm$ 5 & $\pm$ 4 & 
$\pm$ 5 & $\pm$ 5 & $[Fe/H]_{47 Tuc} \pm$ 0.1\\

--30 & +10 & -6 & +10 & -10 & +5 & +12 & +11.5 & LF Correction \\

--9.5 & +1 & --2 & +1 & --2 & $<$ 1 & $<$ 1 & $<$ 1 & No BS \\

--7 & +4 & --3 & +4 & --4 & +4 & +2 & +2 & 10 $\rightarrow$ 11 Gyrs \\

\enddata
\end{deluxetable}
\clearpage

\end{document}